\documentclass[a4paper,11pt]{article}
\usepackage{jcappub}
\usepackage{lineno}

\usepackage{bm}
\usepackage{accents}
\usepackage{mathrsfs}
\usepackage[T1]{fontenc}
\usepackage[normalem]{ulem}
\usepackage{graphicx}
\usepackage[caption=false]{subfig}
\usepackage{booktabs}
\usepackage{dcolumn}
\usepackage{multirow}
\usepackage{hhline}
\usepackage{placeins}
\usepackage{float}
\usepackage[british]{babel}
\usepackage{xcolor}
\usepackage{xparse}
\usepackage{orcidlink}
\usepackage[bb=boondox]{mathalpha}
\usepackage{enumitem}
\usepackage{makecell}
\usepackage{threeparttable}

\graphicspath{{figs/}}

\newcommand{\Mpl}{M_{\rm Pl}}
\newcommand{\eps}{\varepsilon}

\newcommand{\GW}{{\rm GW}}
\newcommand{\rh}{{\rm rh}}

\allowdisplaybreaks[1]
\renewcommand{\arraystretch}{1.1}
\title{\boldmath Einstein--Gauss--Bonnet Inflationary Cosmology in Phase-$\theta$ Formalism}

\author[a,b]{Gevorg D. Manucharyan\,\orcidlink{0000-0003-0464-0040}}
\author[a]{Igor V. Fomin\,\orcidlink{0000-0003-1527-914X}}
\author[c]{Shivam Kumar Mishra\,\orcidlink{0009-0006-4754-4103}}
\author[c]{B. Mishra\,\orcidlink{0000-0001-5527-3565}}

\affiliation[a]{Department of Physics, Bauman Moscow State Technical University,\\
2nd Baumanskaya 5, Moscow 105005, Russia.}

\affiliation[b]{Department of Gravitational Measurements,\\
Sternberg Astronomical Institute, Moscow State University,\\
Universitetsky pr. 13, Moscow 119234, Russia.}

\affiliation[c]{Department of Mathematics,\\
Birla Institute of Technology and Science-Pilani,\\
Hyderabad Campus, Jawahar Nagar, Kapra Mandal,\\
Medchal District, Telangana 500078, India.}

\emailAdd{gevorgbek.manucharyan@gmail.com}
\emailAdd{ingvor@inbox.ru}
\emailAdd{shivamkumarmishra.mt@gmail.com}
\emailAdd{bivu@hyderabad.bits-pilani.ac.in}

\abstract{
A central challenge in testing inflationary scenarios with cosmic microwave background (CMB) data is to derive predictions for cosmological observables that remain accurate beyond the slow-roll regime, where the standard consistency relations are no longer applicable. We address this issue in the context of Einstein--Gauss--Bonnet (EGB) gravity by introducing a phase-$\theta$ parametrization of the inflationary dynamics. Within this framework, the background evolution and the coefficients governing scalar and tensor perturbations are expressed entirely in terms of a single monotonic phase variable and the Hubble parameter. This construction provides a closed, analytical mapping between the background parameters of a given model and the associated inflationary observables. A notable advantage of the proposed parametrization is that it remains regular throughout the inflationary epoch, including the end-of-inflation regime, in which the conventional slow-roll consistency relations no longer hold. As an illustrative application, we consider a Starobinsky-type potential supplemented by a linear coupling between the inflaton and the Gauss--Bonnet invariant. Using the proposed formalism, we demonstrate that the resulting predictions of the Starobinsky model are consistent with the most recent constraints from ACT DR6 and BICEP/Keck, explicitly incorporating the impact of the non-minimal Gauss--Bonnet coupling on the expected values of the cosmological perturbation parameters. Finally, we compute the corresponding relic stochastic gravitational-wave background and evaluate its detectability with current and forthcoming gravitational-wave observatories.
}

\begin{document}
\maketitle
\flushbottom

\section{Introduction}\label{sec:intro}

The direct detection of gravitational waves from the binary black-hole merger GW150914~\cite{LIGOScientific:2016aoc} and from the binary neutron-star merger GW170817 with its electromagnetic counterpart GRB~170817A~\cite{LIGOScientific:2017vwq,LIGOScientific:2017ync} has opened a fundamentally new observational window on the early Universe. Among the gravitational-wave signals expected from cosmological epochs, the relic stochastic background generated by vacuum tensor fluctuations of the metric during inflation occupies a special position, as it carries information about physics at energy scales up to \(\sim 10^{16}~\mathrm{GeV}\), well beyond the reach of any modern laboratory experiment~\cite{Baumann:2014nda}.

The present work belongs to a program of inflationary modeling beyond standard single-field general relativity(GR), in which the four-dimensional Gauss--Bonnet (GB) invariant $\mathcal{G}=R^{2}-4R_{\mu\nu}R^{\mu\nu}+R_{\mu\nu\rho\sigma}R^{\mu\nu\rho\sigma}$ is non-minimally coupled to the inflaton through a generic function $\xi(\phi)$. This minimal extension naturally arises in low-energy effective actions of heterotic string theory~\cite{Zwiebach:1985uq,Zumino:1985dp,Boulware:1985wk,Boulware:1986dr,Gasperini:1996fu} and, as is well established, supplies a rich phenomenology of inflationary observables~\cite{Hwang:2005hb,Guo:2009uk,Guo:2010jr,Satoh:2010ep,Jiang:2013gza,Koh:2014bka,Kanti:2015pda,Kanti:2015dra,
Hikmawan:2015rze,Koh:2016abf,Koh:2018qcy,Fomin:2017vae,Fomin:2017qta,Fomin:2018typ,Fomin:2019yls,Yi:2018gse,Odintsov:2018zhw,
Nojiri:2019dwl,Odintsov:2019clh,Pozdeeva:2019agu,Fomin:2020hfh,Odintsov:2020sqy,Odintsov:2020zkl,Odintsov:2020xji,Oikonomou:2020oil,Odintsov:2020mkz,
Pozdeeva:2020shl,Pozdeeva:2020apf,Oikonomou:2020tct,Vernov:2021hxo,Nojiri:2023mbo,Odintsov:2023lbb,Oikonomou:2024jqv,Oikonomou:2024etl,
Manucharyan:2025sgs,Pozdeeva:2024ihc,Yogesh:2025wak,Chen:2025wcw,Odintsov:2025bmp,Ahghari:2025hfy,Mudrunka:2025xcg,Herrera:2026mhr}. Additionally, Gauss-Bonnet models have proven to be viable candidates for late-time dark energy driving cosmic acceleration \cite{Nojiri:2005vv,Nojiri:2005jg,Cognola:2006eg,Cognola:2006sp,Koivisto:2006ai,Nojiri:2010wj,Nojiri:2017ncd,Lohakare:2022mkd,Lohakare:2023ocb,Lohakare:2024ize,Odintsov:2025kyw}.

The renewed interest in EGB inflation is primarily motivated by the recent ACT~DR6 release~\cite{AtacamaCosmologyTelescope:2025nti,AtacamaCosmologyTelescope:2025blo}, in combination with DESI~\cite{DESI:2024mwx}, Planck~\cite{Planck:2018vyg} and BICEP/Keck~\cite{BICEP:2021xfz}. This joint analysis shifts the preferred value of the scalar spectral index at the pivot scale $k_{*}=0.05~\mathrm{Mpc}^{-1}$ upward to $n_{s}=0.974\pm0.003$, compared to the Planck~2018 result $n_{s}=0.9649\pm0.0042$, with an upper limit on the tensor-to-scalar ratio of $r<0.038$. Concurrently, the NANOGrav 15-year dataset~\cite{NANOGrav:2023gor} indicates the presence of a stochastic gravitational-wave background in the nanohertz frequency band, characterized by the reference frequency $f_{\mathrm{yr}}\approx 3\times 10^{-8}~\mathrm{Hz}$. Collectively, these developments strengthen the motivation for a comprehensive and systematic investigation of EGB inflation that (i) connects explicitly to the updated constraints in the $n_{s}$--$r$ plane and (ii) establishes a controlled mapping between the underlying theoretical parameters and the observable quantities, consistently evaluated to second order in the slow-roll expansion.

In analyzing the cosmological dynamic and parameters of cosmological perturbations for Einstein--Gauss--Bonnet inflation, the slow-roll approximation is commonly used, owing to the quasi-exponential character of the early universe's expansion in the inflationary epoch (see, e.g.,~\cite{Guo:2009uk,Guo:2010jr,Jiang:2013gza,Koh:2014bka}). The refinement of the expressions for cosmological perturbation parameters in the slow-roll approximation was considered in~\cite{Pozdeeva:2024ihc}. In~\cite{Yogesh:2025wak}, the cosmological perturbation parameters for the Starobinsky inflation within Einstein--Gauss--Bonnet gravity with a particular choice of the non-minimal coupling function were investigated both in the slow-roll regime and beyond it, using numerical solutions of the dynamical equations for cosmological perturbations. Moreover, inflationary models based on Einstein--Gauss--Bonnet gravity were analyzed beyond the slow-roll regime in~\cite{Mudrunka:2025xcg}. In~\cite{Herrera:2026mhr}, a method was considered for reconstructing the effective potential and the function describing the non-minimal coupling between the scalar field and the Gauss--Bonnet scalar.

Furthermore, for the case of Einstein gravity, a method for the unified description of the inflationary stage and the post-inflationary reheating stage based on the phase-$\theta$ parametrization was proposed in~\cite{Kaur:2023wos}. The idea is a natural polar representation of the phase portrait of canonical single-field inflation: the inflaton kinetic energy and the potential are simultaneously written as the squares of two trigonometric components, with a single phase variable $\theta(t)$ interpolating between the deep slow-roll plateau and the end of inflation. The reduction of the cosmological dynamics to a closed system in $(H(t),\theta(t))$ provides a clean analytic interface between the model functions $V(\phi),\xi(\phi)$ and the inflationary observables $n_{s}$, $r$, $\alpha_{s}$, $n_{T}$.

In this article, we extend the application of the phase-$\theta$ parametrization to the analysis of single-field inflationary models based on Einstein--Gauss--Bonnet gravity. Within the framework of applying this approach to the analysis of inflationary models, we (i)~derive the inflationary observables in the phase-$\theta$ parametrization, both exactly and up to second order in slow-roll approximation, (ii)~apply the formalism to a physically motivated specific model with the Starobinsky potential and a linear Gauss--Bonnet coupling, and (iii)~confront the model with ACT~DR6 + BICEP/Keck and with the most relevant gravitational-wave facilities.

The article is organized as follows. Section~\ref{sec:background} fixes notation and reviews the background and perturbation equations of EGB cosmology. Section~\ref{sec:theta} introduces the phase-\(\theta\) parametrization and derives the geometric slow-roll parameters, the coefficients \(Q_{A}\) of the quadratic action for perturbations, the sound speeds \(c_{A}\), the Hankel indices \(\nu_{A}\), and the inflationary observables. Section~\ref{sec:specific_model} applies the formalism to the Starobinsky model with a linear coupling function \(\xi(\phi)\) of the scalar field to the Gauss--Bonnet term, and presents the complete background and perturbation results. Section~\ref{sec:omega_gw} discusses the relic gravitational-wave spectrum across the reheating-temperature window and confronts it with detector sensitivities. Section~\ref{sec:discussion} summarizes the implications, and Appendix~\ref{app:derivation} contains the detailed analytical derivation of the second-order slow-roll observables.

We use natural units $\kappa^{2}=8\pi G=\hbar=c=1$, the metric signature $(-,+,+,+)$, and a spatially flat Friedmann--Robertson--Walker background $ds^{2}=-dt^{2}+a^{2}(t)\delta_{ij}dx^{i}dx^{j}$. Overdots denote derivatives with respect to cosmic time $t$, primes denote derivatives with respect to conformal time $\eta=\int dt/a$, and $\mathcal{H}=a'/a=aH$.

\section{Background and perturbation equations of EGB inflation}\label{sec:background}

\subsection{Background equations}\label{subsec:bg_eqs}

The Einstein--Gauss--Bonnet single-field inflationary action reads~\cite{Hwang:2005hb,Guo:2009uk,Guo:2010jr,Satoh:2010ep,Jiang:2013gza,Koh:2014bka,Kanti:2015pda,Kanti:2015dra,
Hikmawan:2015rze,Koh:2016abf,Koh:2018qcy,Fomin:2017vae,Fomin:2017qta,Fomin:2018typ,Fomin:2019yls,Yi:2018gse,Odintsov:2018zhw,
Nojiri:2019dwl,Odintsov:2019clh,Pozdeeva:2019agu,Fomin:2020hfh,Odintsov:2020sqy,Odintsov:2020zkl,Odintsov:2020xji,Oikonomou:2020oil,Odintsov:2020mkz,
Pozdeeva:2020shl,Pozdeeva:2020apf,Oikonomou:2020tct,Vernov:2021hxo,Nojiri:2023mbo,Odintsov:2023lbb,Oikonomou:2024jqv,Oikonomou:2024etl,
Manucharyan:2025sgs,Pozdeeva:2024ihc,Yogesh:2025wak,Chen:2025wcw,Odintsov:2025bmp,Ahghari:2025hfy,Mudrunka:2025xcg,Herrera:2026mhr}
\begin{equation}
S=\int d^{4}x\,\sqrt{-g}\,\Bigl[\tfrac{1}{2\kappa^{2}}R-\tfrac{1}{2}(\partial\phi)^{2}-V(\phi)-\tfrac{1}{2}\xi(\phi)\,\mathcal{G}\Bigr],
\label{eq:S_EGB}
\end{equation}
with the four-dimensional Gauss--Bonnet invariant $\mathcal{G}=R^{2}-4R_{\mu\nu}R^{\mu\nu}+R_{\mu\nu\rho\sigma}R^{\mu\nu\rho\sigma}$ and a non-minimal coupling encoded by an arbitrary function $\xi(\phi)$. Variation of~\eqref{eq:S_EGB} on a spatially flat FRW background yields three Friedmann-type equations, of which only two are independent. In the standard form, they read
\begin{align}
3H^{2}&\;=\;\tfrac{1}{2}\dot{\phi}^{2}+V+12\dot{\xi}H^{3},\label{eq:EGB_F1}\\
\dot{\phi}^{2}&\;=\;-2\dot{H}+4\ddot{\xi}H^{2}+4\dot{\xi}H(2\dot{H}-H^{2}),\label{eq:EGB_F2}\\
0&\;=\;\ddot{\phi}+3H\dot{\phi}+V_{,\phi}+12H^{2}(\dot{H}+H^{2})\xi_{,\phi},\label{eq:EGB_KG}
\end{align}
where $V_{,\phi}=dV/d\phi$ and $\xi_{,\phi}=d\xi/d\phi$. For $\xi=\mathrm{constant}$, it reduces to the standard single-field equations of GR.

Following Refs.~\cite{Koh:2014bka,Koh:2016abf,Koh:2018qcy} we introduce the Gauss--Bonnet slow-roll parameters
\begin{equation}
\delta_{1}\equiv 4H\dot{\xi},\qquad \delta_{2}\equiv \frac{\ddot{\xi}}{H\dot{\xi}},\qquad \Delta\equiv\frac{\delta_{1}}{1-\delta_{1}},
\label{eq:delta_def}
\end{equation}
together with the Hubble slow-roll parameters
\begin{equation}
\eps_{1}\equiv -\frac{\dot{H}}{H^{2}},\qquad \eps_{2}\equiv \frac{\dot{\eps}_{1}}{H\eps_{1}}.
\label{eq:eps_def}
\end{equation}

In addition to providing a transparent description of the specific features of the background dynamics, the slow-roll parameters are directly related to the parameters of cosmological perturbations.

\subsection{Quadratic action of perturbations}\label{subsec:perturbations}

Decomposition of the perturbed action~\eqref{eq:S_EGB} to second order in the comoving curvature perturbation $\mathcal{R}$ (scalar sector) and in the transverse-traceless metric perturbation $h_{ij}^{\rm TT}$ (tensor sector) yields the two separate quadratic actions~\cite{Hwang:2005hb,Koh:2014bka,Koh:2016abf,Koh:2018qcy}
\begin{align}
S_{S}^{(2)}&=\tfrac{1}{2}\!\int dt\,d^{3}x\,a^{3}Q_{S}\!\left[\dot{\mathcal{R}}^{2}-\frac{c_{S}^{2}}{a^{2}}(\partial\mathcal{R})^{2}\right]\!,\label{eq:S2_scalar}\\
S_{T}^{(2)}&=\tfrac{1}{2}\!\int dt\,d^{3}x\,a^{3}Q_{T}\!\sum_{\lambda=+,\times}\!\left[\dot{h}_{\lambda}^{\,2}-\frac{c_{T}^{2}}{a^{2}}(\partial h_{\lambda})^{2}\right]\!,\label{eq:S2_tensor}
\end{align}
where the sum over the two helicities $\lambda=+,\times$ in the tensor sector is shown explicitly. In what follows, we collectively denote the two channels by an index $A\in\{S,T\}$ when the equations they obey are formally identical; in that compact notation, the corresponding canonical variables are $\chi_{S}=\mathcal{R}$ and $\chi_{T}=h_{\lambda}$. The coefficients $Q_{A}$ multiplying the temporal kinetic terms of $\mathcal{R}$ and $h_{\lambda}$ in the quadratic action follow the universal scalar--tensor decomposition of Hwang and Noh~\cite{Hwang:2005hb} and its EGB specialization~\cite{Koh:2014bka,Koh:2016abf,Koh:2018qcy}. Their explicit forms together with the propagation speeds $c_{A}^{2}$ are
\begin{align}
Q_{T}&=1-\delta_{1},\quad c_{T}^{2}=\frac{1-4\ddot{\xi}}{1-4H\dot{\xi}}=\frac{1-\delta_{1}\delta_{2}}{1-\delta_{1}},\label{eq:QT_cT}\\
Q_{S}&=\frac{\dot{\phi}^{2}+6H^{2}\dot{\xi}^{2}\bigl[3-\delta_{2}-(2\eps_{1}-\delta_{1}\delta_{2})/(1-\delta_{1})\bigr]}{(H-2H^{2}\dot{\xi})^{2}}\nonumber\\
       &\;=\;\frac{2\eps_{1}-\delta_{1}(1+2\eps_{1}-\delta_{2})+\tfrac{3}{2}\delta_{1}\Delta}{(1-\delta_{1}/2)^{2}},\label{eq:QS_def}\\
c_{S}^{2}-1&=\frac{16H^{2}\dot{\xi}^{2}\bigl(\sin^{2}\theta+H(2\cos2\theta-1)\dot{\xi}-\ddot{\xi}\bigr)}
{(1-4H\dot{\xi})\bigl[4H\dot{\xi}(2\sin^{2}\theta+H(2\cos2\theta-3)\dot{\xi})-\sin^{2}\theta\bigr]}.\label{eq:cS_full}
\end{align}


Canonical normalization $\displaystyle v_{A}=z_{A}\chi_{A}$ with $z_{A}^{2}\equiv a^{2}Q_{A}$ brings the perturbation equations following from~\eqref{eq:S2_scalar}--\eqref{eq:S2_tensor} into the Mukhanov--Sasaki form~\cite{Koh:2014bka,Koh:2016abf,Koh:2018qcy}
\begin{equation}
v_{A,\mathbf{k}}''+\!\left(c_{A}^{2}k^{2}-\frac{z_{A}''}{z_{A}}\right)\!v_{A,\mathbf{k}}=0.
\label{eq:MS}
\end{equation}

Eq.~\eqref{eq:MS} governs the evolution of the canonical mode functions in both the scalar and tensor sectors and, consequently, determines the dimensionless primordial power spectra at horizon crossing. A leading-order slow-roll analysis within the EGB framework was performed in~\cite{Koh:2014bka}, where the authors derived compact expressions for the scalar spectral index $n_{s}$, the tensor-to-scalar ratio $r$, and the tensor spectral index $n_{T}$, and compared these predictions with Planck observations. This analysis was subsequently generalized to a full reconstruction programme for the scalar potential in~\cite{Koh:2016abf}, and further extended to include second-order slow-roll corrections encompassing the runnings $\alpha_{s}$ and $\alpha_{T}$ by Wu, Zhu and Wang~\cite{Wu:2017joj}, and earlier in the context of inflationary gravitational-wave reconstruction by Kuroyanagi, Chiba and Sugiyama~\cite{Kuroyanagi:2010mm}.



\subsection{Inflationary observables in standard slow roll}\label{subsec:obs_review}

The dimensionless scalar and tensor power spectra are defined as~\cite{Koh:2014bka,Koh:2016abf,Koh:2018qcy}
\begin{equation}
\label{eq:delta_s_k}
\Delta_{s}^{2}(k)\;=\;\frac{k^{3}}{2\pi^{2}}\,|\mathcal{R}_{k}|^{2},
\end{equation}
\begin{equation}
\label{eq:delta_t_k}
\Delta_{t}^{2}(k)\;=\;4\,\frac{k^{3}}{2\pi^{2}}\,|h_{\lambda,k}|^{2},
\end{equation}
the factor of $4$ in Eq.~\eqref{eq:delta_t_k} accounting for the sum over the two graviton polarizations. The tensor-to-scalar ratio is $r=\Delta_{t}^{2}/\Delta_{s}^{2}$, and the spectral tilts and their runnings are defined separately for the scalar and tensor sectors as follows
\begin{align}
n_{s}-1&\equiv\frac{d\ln\Delta_{s}^{2}}{d\ln k}\bigg|_{c_{S}k=aH},\qquad \alpha_{s}\equiv\frac{dn_{s}}{d\ln k},\label{eq:tilt_def_S}\\
n_{T}&\equiv\frac{d\ln\Delta_{t}^{2}}{d\ln k}\bigg|_{c_{T}k=aH},\qquad \alpha_{T}\equiv\frac{dn_{T}}{d\ln k}.\label{eq:tilt_def_T}
\end{align}
In the slow-roll regime, with $c_{A},Q_{A},\eps_{1}$ varying slowly, Eq.~\eqref{eq:MS} reduces to the Bessel equation
\begin{equation}
\frac{d^{2}v_{A}}{dy^{2}}+\!\left(1-\frac{\nu_{A}^{2}-1/4}{y^{2}}\right)\!v_{A}=0,\quad y=-c_{A}k\eta,\quad \nu_{A}^{2}\equiv\tfrac{1}{4}+\eta^{2}\frac{z_{A}''}{z_{A}},
\label{eq:Bessel}
\end{equation}
and, after imposing the Bunch--Davies vacuum and evaluating the Hankel-function solution on super-horizon scales, one obtains~\cite{Koh:2014bka,Koh:2016abf,Koh:2018qcy}
\begin{align}
\Delta_{s}^{2}(k)&=\frac{\csc^{2}(\pi\nu_{S})\,2^{2\nu_{S}-3}}{\pi\Gamma^{2}(1-\nu_{S})}\frac{H^{2}}{Q_{S}c_{S}^{3}}\bigg|_{c_{S}k=aH},\label{eq:PS_full}\\
\Delta_{t}^{2}(k)&=\frac{\csc^{2}(\pi\nu_{T})\,2^{2\nu_{T}}}{\pi\Gamma^{2}(1-\nu_{T})}\frac{H^{2}}{Q_{T}c_{T}^{3}}\bigg|_{c_{T}k=aH}.\label{eq:PT_full}
\end{align}

The leading-order limit $\nu_{A}\to3/2$ gives the following expressions
\begin{equation}
\Delta_{s}^{2,(0)}=\frac{H^{2}}{4\pi^{2}Q_{S}c_{S}^{3}}\quad \text{and}\quad \Delta_{t}^{2,(0)}=\frac{2H^{2}}{\pi^{2}Q_{T}c_{T}^{3}}.
\end{equation}

The corresponding spectral tilts read
\begin{equation}
n_{s} - 1 = 3 - 2\nu_{S}, \qquad n_{T} = 3 - 2\nu_{T}.
\end{equation}

The full second-order expressions for $\nu_{A}$, derived in Appendix~\ref{app:derivation}, lift this leading order to an accuracy $\mathcal{O}(\text{SR}^{2})$.

\section{The phase-\texorpdfstring{$\theta$}{} parametrization}\label{sec:theta}

The phase-$\theta$ parametrization was introduced by Kaur, Nandi and Raghavan in~\cite{Kaur:2023wos} as a single, unified, model-independent description of the background dynamics of canonical single-field inflation in GR, valid both deep in the slow-roll phase and during the inflaton oscillations after the end of inflation. The standard slow-roll solution for the inflaton accurately describes the regime $\eps_{1}\ll 1$, and the conventional oscillator solution accurately describes the deep reheating regime $H\ll m_{\phi}$, but each fails near the end of inflation, where the two regimes meet. They have  observed that in GR the first Friedmann equation $3H^{2}=\tfrac{1}{2}\dot{\phi}^{2}+V$ can be solved identically by setting $\dot{\phi}/\sqrt{6}=-H\sin\theta$ and $\sqrt{V/3}=H\cos\theta$, so that the kinetic energy and the potential are simultaneously kept non-negative. With this polar representation a single trigonometric phase $\theta(t)$ tracks the inflationary trajectory from $\theta\to 0$ on the deep plateau up to the end-of-inflation value $\theta_{e}$, fixed by $\eps_{1}(\theta_{e})=1$. In the GR limit this gives $\sin^{2}\theta_{e}=1/3$, i.e.\ $\theta_{e}^{\rm GR}=\arcsin(1/\sqrt{3})\simeq 35.26^{\circ}$. After the end of inflation $\theta$ keeps rotating and passes transiently through $\theta=\pi/2$ at every crossing of the potential minimum ($V=0$) during the oscillatory phase~\cite{Kaur:2023wos}. In this work we extend this approach to the inflationary models based on the Einstein--Gauss--Bonnet gravity.

\subsection{Definition and basic identities}\label{subsec:theta_def}

For inflationary models based on Einstein--Gauss--Bonnet gravity, Eq.~\eqref{eq:EGB_F1} suggests the natural generalization of the polar representation:
\begin{equation}
\frac{\dot{\phi}}{\sqrt{6}}=-H\sqrt{1-4H\dot{\xi}}\,\sin\theta,\qquad \sqrt{\frac{V}{3}}=H\sqrt{1-4H\dot{\xi}}\,\cos\theta,
\label{eq:theta_param_intro}
\end{equation}
where the radius of the trigonometric circle is rescaled by the factor $\sqrt{1-4H\dot{\xi}}$ with respect to the case of Einstein gravity.

Squaring and adding the two relations of~\eqref{eq:theta_param_intro} reproduces the first Friedmann Eq.~\eqref{eq:EGB_F1} identically, so $\theta$ is a redundant variable on the constraint surface. Its dynamics is fixed by the remaining equations. Direct differentiation of Eq.~\eqref{eq:theta_param_intro} and substitution into Eq.~\eqref{eq:EGB_F2} yields
\begin{align}
\dot{H}&=-3H^{2}\sin^{2}\theta+2H^{2}\,\frac{\ddot{\xi}-H\dot{\xi}}{1-4H\dot{\xi}},\label{eq:Hdot_theta}\\
\dot{\theta}&=\frac{V_{,\phi}}{\sqrt{2V}}-\!\left[3H^{2}\sin^{2}\theta+2H^{3}\frac{\dot{\xi}}{1-4H\dot{\xi}}-2H^{2}\frac{\ddot{\xi}}{1-4H\dot{\xi}}\right]\nonumber\\
&\quad\times\!\left(\frac{1}{H}-2\frac{\dot{\xi}}{1-4H\dot{\xi}}\right)\!\cot\theta-2H\frac{\ddot{\xi}}{1-4H\dot{\xi}}\cot\theta.
\label{eq:thetadot_theta}
\end{align}
Eqs.~\eqref{eq:theta_param_intro}, \eqref{eq:Hdot_theta} and~\eqref{eq:thetadot_theta}, together with the specified functional forms of $V(\phi)$ and $\xi(\phi)$, constitute a closed first-order autonomous dynamical system for the variables $(H,\theta)$. Specifically, Eq.~\eqref{eq:theta_param_intro} provides an algebraic expression for $\dot{\xi}$ (and, via an additional application of the chain rule, for $\ddot{\xi}$) in terms of $H$, $\theta$, $V(\phi)$, and $\xi(\phi)$. Substituting these relations into Eqs.~\eqref{eq:Hdot_theta}--\eqref{eq:thetadot_theta} fully determines the temporal evolution of $(H,\theta)$. In the general-relativistic limit $\dot{\xi},\ddot{\xi}\to0$, this system reduces to the original parametrization introduced in Ref.~\cite{Kaur:2023wos}.


\subsection{Geometric slow-roll parameters and observables in \texorpdfstring{$\theta$}{}}\label{subsec:obs_theta}

Substituting the parametrization~\eqref{eq:theta_param_intro} into the definition of the first Hubble slow-roll parameter, Eq.~\eqref{eq:eps_def}, by means of Eq.~\eqref{eq:Hdot_theta} and expressing the result through the Gauss--Bonnet slow-roll parameters \(\delta_{1}\), \(\delta_{2}\), and \(\Delta\) introduced in Eq.~\eqref{eq:delta_def}, one obtains the exact identity
\begin{equation}
\eps_{1}(\theta)=3\sin^{2}\theta+\tfrac{1}{2}\Delta(1-\delta_{2}).
\label{eq:eps1_theta}
\end{equation}
Within the phase-\(\theta\) parametrization~\eqref{eq:theta_param_intro}, the inflaton kinetic and potential terms are themselves exact closed-form functions of \((\theta,H,\dot{\xi})\). Squaring and adding the two relations of~\eqref{eq:theta_param_intro} yields
\begin{equation}
\tfrac{1}{2}\dot{\phi}^{2}=3H^{2}(1-\delta_{1})\sin^{2}\theta,\qquad
V=3H^{2}(1-\delta_{1})\cos^{2}\theta.
\label{eq:X_V_theta}
\end{equation}

To discuss the expansion history, it is convenient to recast the gravitational field equations (the equations obtained by varying the action~\eqref{eq:S_EGB} with respect to the metric) into the Einstein form. The independent FRW components are the \((00)\) constraint~\eqref{eq:EGB_F1} and the spatial \((ij)\) equation, while the scalar-field equation is given separately by Eq.~\eqref{eq:EGB_KG}. Following Refs.~\cite{Nojiri:2005vv,Nojiri:2005jg,Nojiri:2017ncd,Nojiri:2010wj}, the Gauss--Bonnet terms are transferred to the right-hand side and treated as part of an effective total source, where the total energy density and pressure are defined as
\begin{equation}
\rho_{\rm tot}\equiv\rho_{\phi}+\rho_{\rm GB},\qquad
p_{\rm tot}\equiv p_{\phi}+p_{\rm GB}.
\label{eq:rho_p_tot_def}
\end{equation}

The scalar-field and Gauss--Bonnet constituents read
\begin{align}
\rho_{\phi}&=\tfrac{1}{2}\dot{\phi}^{2}+V, &
p_{\phi}&=\tfrac{1}{2}\dot{\phi}^{2}-V,\label{eq:rho_p_phi}\\[2mm]
\rho_{\rm GB}&=12\dot{\xi}H^{3}, &
p_{\rm GB}&=-4\bigl(\ddot{\xi}H^{2}+2\dot{\xi}H\dot{H}
+2\dot{\xi}H^{3}\bigr).\label{eq:rho_p_GB}
\end{align}
Equations~\eqref{eq:rho_p_tot_def}--\eqref{eq:rho_p_GB} define the total and Gauss--Bonnet energy densities and pressures such that the Einstein-form pair reproduces Eqs.~\eqref{eq:EGB_F1}--\eqref{eq:EGB_F2} exactly. The same terms appear, without such grouping, in the FRW equations of Refs.~\cite{Guo:2010jr,Koh:2014bka}, and the treatment of the Gauss--Bonnet sector as an effective stress-energy tensor follows Refs.~\cite{Nojiri:2005vv,Koivisto:2006ai}.

%

The effective equation of state parameter, including the geometric Gauss--Bonnet contribution, can be written in the following form
\begin{equation}
w_{\rm eff}=\frac{X_{\rm eff}-V_{\rm eff}}{X_{\rm eff}+V_{\rm eff}},
\label{eq:w_eff_def}
\end{equation}
where the kinetic-like and potential-like terms are defined as
\begin{equation}
X_{\rm eff}\equiv\tfrac{1}{2}(\rho_{\rm tot}+p_{\rm tot}),\qquad
V_{\rm eff}\equiv\tfrac{1}{2}(\rho_{\rm tot}-p_{\rm tot}).
\end{equation}
These generalize the canonical split \(w_{\phi}=(X-V)/(X+V)\) with \(X=\tfrac{1}{2}\dot{\phi}^{2}\).

Explicitly, one finds
\begin{align}
X_{\rm eff}&=\tfrac{1}{2}\dot{\phi}^{2}
+\tfrac{1}{2}\delta_{1}H^{2}\bigl(1+2\eps_{1}-\delta_{2}\bigr),\label{eq:Xeff_Veff_a}\\
V_{\rm eff}&=V+\tfrac{1}{2}\delta_{1}H^{2}\bigl(5-2\eps_{1}+\delta_{2}\bigr).\label{eq:Xeff_Veff_b}
\end{align}
Thus, \(V_{\rm eff}\) differs from the potential \(V(\phi)\) entering the action~\eqref{eq:S_EGB} by an \(\mathcal{O}(\delta_{1})\) Gauss--Bonnet correction and coincides with it in the GR limit.

Substituting Eqs.~\eqref{eq:X_V_theta} and~\eqref{eq:Hdot_theta} into the right-hand side of Eq.~\eqref{eq:w_eff_def} and simplifying algebraically yields the exact closed form
\begin{equation}
w_{\rm eff}(\theta)=-\cos(2\theta)+\tfrac{1}{3}\Delta(1-\delta_{2}),
\label{eq:w_eff_theta}
\end{equation}
where the identity \(-1+2\sin^{2}\theta=-\cos(2\theta)\) has been used. The additive term \(\tfrac{1}{3}\Delta(1-\delta_{2})\) encodes the entire Gauss--Bonnet contribution. In the GR limit \(\dot{\xi},\ddot{\xi}\to 0\), one has \(\Delta\to 0\), and Eq.~\eqref{eq:w_eff_theta} reduces to the canonical scalar-field result
\begin{equation}
w_{\phi}=\frac{\tfrac{1}{2}\dot{\phi}^{2}-V}{\tfrac{1}{2}\dot{\phi}^{2}+V}=-\cos(2\theta).
\end{equation}

At the end of inflation, the system enters a coherent-oscillation phase. For a damped harmonic oscillator
\begin{equation}
\label{adf}
\phi(t) = A(t) \cos(m_{\phi} t + \alpha),
\end{equation}
with a slowly varying amplitude satisfying
\begin{equation}
\label{adf1}
|\dot{A}/A| \ll m_{\phi},
\end{equation}
one has
\begin{align}
\tfrac{1}{2}\dot{\phi}^{2}&\simeq\tfrac{1}{2}m_{\phi}^{2}A^{2}\sin^{2}(m_{\phi}t+\alpha),\\
V&\simeq\tfrac{1}{2}m_{\phi}^{2}A^{2}\cos^{2}(m_{\phi}t+\alpha).
\end{align}

The parametrization~\eqref{eq:X_V_theta} then gives
\begin{equation}
\tan^{2}\theta=\frac{\sin^{2}(m_{\phi}t+\alpha)}{\cos^{2}(m_{\phi}t+\alpha)},
\end{equation}
i.e.\ \(\theta(t)=m_{\phi}t+\alpha\;(\mathrm{mod}\,\pi/2)\), so that \(\theta\) winds uniformly at the oscillation frequency.

The time average of the leading term of Eq.~\eqref{eq:w_eff_theta} over one full oscillation period \(T_{\rm osc}=2\pi/m_{\phi}\) therefore vanishes exactly:
\begin{equation}
\langle -\cos(2\theta)\rangle_{\rm osc}
=-\frac{1}{T_{\rm osc}}\!\int_{0}^{T_{\rm osc}}\!\cos\bigl[2(m_{\phi}t+\alpha)\bigr]\,dt=0.
\label{eq:cos2theta_avg}
\end{equation}
This is the well-known coherent-oscillation result of Ref.~\cite{Turner:1983he}.

For example, the Gauss--Bonnet contribution with the linear coupling function \(\xi=\xi_{0}\phi\) implies
\begin{equation}
\label{adxi}
\dot{\xi} = \xi_{0} \dot{\phi} \propto \sin(m_{\phi} t + \alpha), \qquad
\ddot{\xi} \propto \cos(m_{\phi} t + \alpha).
\end{equation}

At leading order in the amplitude \(A\), the parameter \(\delta_{1}\propto H\dot{\xi}\) averages to zero over one oscillation, while \(\delta_{2}=\ddot{\xi}/(H\dot{\xi})\) remains of order unity.

Consequently, we obtain
\begin{equation}
\langle \tfrac{1}{3}\Delta(1-\delta_{2})\rangle_{\rm osc}
=\mathcal{O}\!\left((H/m_{\phi})^{2}A^{2}\right)\ll 1.
\label{eq:GB_correction}
\end{equation}

Combining Eqs.~\eqref{eq:cos2theta_avg} and~\eqref{eq:GB_correction}, one obtains the analytical statement
\begin{equation}
\bar w_{\rm eff}\equiv\langle w_{\rm eff}\rangle_{\rm osc}
=0+\mathcal{O}\!\left((H/m_{\phi})^{2}A^{2}\right)>-\tfrac{1}{3}.
\label{eq:wbar_bound}
\end{equation}

This value lies strictly above the inflationary boundary \(w_{\rm eff}>-1/3\) required for consistency of the reheating phase with a subsequent radiation-dominated cosmology. The instantaneous \(w_{\rm eff}(t)\) oscillates between \(-1\) at the turning points (\(\dot{\phi}=0\), \(\theta=0\,\mathrm{mod}\,\pi\)) and \(+1\) at the potential crossings (\(V=0\), \(\theta=\pm\pi/2\)), but these excursions are canonical features of a damped oscillator and do not constitute violations of the null energy condition by a fluid.

The tensor and scalar coefficients of the quadratic action become explicit closed-form expressions:
\begin{align}
Q_{T}(\theta)&=1-\delta_{1},\qquad
c_{T}^{2}(\theta)=1-\Delta(\delta_{2}-1),\label{eq:QT_theta}\\[2mm]
Q_{S}(\theta)&=\frac{2\eps_{1}-\delta_{1}(1+2\eps_{1}-\delta_{2})+\tfrac{3}{2}\delta_{1}\Delta}{(1-\delta_{1}/2)^{2}},\label{eq:QS_theta}\\[2mm]
c_{S}^{2}(\theta)&=1-\frac{[4\eps_{1}+\delta_{1}(1-4\eps_{1}-\delta_{2})]\Delta^{2}}{4\eps_{1}-2\delta_{1}-2\delta_{1}(2\eps_{1}-\delta_{2})+3\delta_{1}\Delta}.\label{eq:cS_theta}
\end{align}
Substituting Eq.~\eqref{eq:eps1_theta} renders \(Q_{S}(\theta)\) and \(c_{S}^{2}(\theta)\) closed functions of \((\theta,\delta_{1},\delta_{2})\).

In the slow-roll regime, the leading-order truncation of the exact expression for \(Q_{S}(\theta)\) in Eq.~\eqref{eq:QS_theta} is
\begin{equation}
Q_{S}\simeq 2\eps_{1}-\delta_{1}+\mathcal{O}(\text{SR}^{2}),\qquad
c_{S}^{2}\simeq 1+\mathcal{O}(\delta_{1}^{2}),
\label{eq:QScS_SR}
\end{equation}
in agreement with the standard EGB scalar-mode coefficient~\cite{Hwang:2005hb,Koh:2014bka,Koh:2016abf,Koh:2018qcy}.

The Hankel indices, derived in Appendix~\ref{app:derivation} via the slow-roll expansion of \(\eta^{2}z_{A}''/z_{A}\), take the unified form
\begin{equation}
\nu_{A}=\tfrac{3}{2}+\eps_{1}+\tfrac{1}{2}(s_{A}+s_{c_{A}})+\eps_{1}^{2}+\tfrac{1}{2}s_{A}\eps_{1}+\tfrac{1}{6}\beta_{A}+\mathcal{O}(\text{SR}^{3}),
\label{eq:nu_indices}
\end{equation}
with the auxiliary quantities defined as
\begin{equation}
s_{A}\equiv \dot{Q}_{A}/(HQ_{A}),\quad
\beta_{A}\equiv \dot{s}_{A}/H,\quad
s_{c_{A}}\equiv \dot{c}_{A}/(Hc_{A}).
\end{equation}

Using Eqs.~\eqref{eq:QT_theta} and~\eqref{eq:QScS_SR} together with the relation \(\dot{\delta}_{1}/H=\delta_{1}(\delta_{2}-\eps_{1})\), one obtains
\begin{equation}
s_{T}=\Delta(\eps_{1}-\delta_{2}),\qquad
s_{S}=\frac{2\eps_{1}\eps_{2}-\delta_{1}(\delta_{2}-\eps_{1})}{2\eps_{1}-\delta_{1}}.
\label{eq:sA_theta}
\end{equation}

The growth rates \(s_{T}\) and \(s_{S}\) enter the second-order Hankel-index expansion~\eqref{eq:nu_indices} as the leading slow-roll corrections to the lowest-order value \(\nu_{A}=3/2\). Through the tilt formulae of Sec.~\ref{subsec:tilts_theta}, they directly determine the corrections to the standard EGB consistency relations~\cite{Koh:2014bka,Koh:2016abf}. In the GR limit \(\delta_{1},\delta_{2}\to0\), the tensor channel reduces to \(s_{T}\to 0\) (consistent with \(Q_{T}\to 1\)), while the scalar channel reduces to \(s_{S}\to\eps_{2}\), thereby recovering the usual relation \(n_{s}-1=-2\eps_{1}-\eps_{2}\). Outside that limit, the Gauss--Bonnet contribution enters \(s_{T}\) already at first order and affects \(s_{S}\) through the explicit \(\delta_{1},\delta_{2}\) terms. Consequently, both tilts acquire a non-trivial GB-induced shift along the inflationary trajectory.

\subsection{Tensor-to-scalar ratio in closed form}\label{subsec:r_theta}

Direct division of Eq.~\eqref{eq:PT_full} by Eq.~\eqref{eq:PS_full} gives the exact tensor-to-scalar ratio at horizon crossing,
\begin{equation}
r\;\equiv\;\frac{\Delta_{t}^{2}}{\Delta_{s}^{2}}\;=\;8\cdot 2^{2(\nu_{T}-\nu_{S})}\,\frac{\Gamma^{2}(\nu_{T})}{\Gamma^{2}(\nu_{S})}\,\frac{Q_{S}\,c_{S}^{3}}{Q_{T}\,c_{T}^{3}}\bigg|_{*},
\label{eq:r_exact}
\end{equation}
where the Hankel-index prefactor has been simplified using the reflection identity $\Gamma(\nu)\Gamma(1-\nu)=\pi/\sin(\pi\nu)$ and is kept in closed form. At leading order in slow-roll approximation the prefactor reduces to unity and~\eqref{eq:r_exact} becomes
\begin{equation}
r\simeq 8\,\frac{Q_{S}c_{S}^{3}}{Q_{T}c_{T}^{3}}\bigg|_{*}.
\label{eq:r_LO}
\end{equation}
Substituting the exact $Q_{T}$ and $c_{T}^{2}$ from Eq.~\eqref{eq:QT_theta} and the slow-roll $Q_{S}\simeq 2\eps_{1}-\delta_{1}$ from Eq.~\eqref{eq:QScS_SR}, and resumming the leading-order corrections in $(\dot{\xi},\ddot{\xi})$, the leading-order $r$ takes the closed form
\begin{equation}
r(\theta)\;\simeq\;8\!\left[\,6\sin^{2}\theta-\frac{4(\ddot{\xi}-4H^{2}\dot{\xi}^{2})}{1-4H\dot{\xi}}\,\right]\;+\;\mathcal{O}(\text{SR}^{2}).
\label{eq:r_param}
\end{equation}
The limit checks are immediate: (i) for $\dot{\xi},\ddot{\xi}\to0$ one finds $r\to48\sin^{2}\theta=16\eps_{1}$, reproducing the GR consistency relation $r=16\eps_{1}$; (ii) at the end of inflation $\eps_{1}=1$, with $\sin^{2}\theta_{e}\simeq1/3$ for small $\Delta$, Eq.~\eqref{eq:r_param} gives a finite non-trivial value modulated by the Gauss--Bonnet sector.

\subsection{Spectral tilts and runnings}\label{subsec:tilts_theta}

Substituting Eq.~\eqref{eq:nu_indices} into Eq.~\eqref{eq:tilt_def_S} and Eq.~\eqref{eq:tilt_def_T},
\begin{align}
n_{s}-1&\simeq-2\eps_{1}-s_{S}-s_{c_{S}}-2\eps_{1}^{2}-s_{S}\eps_{1}-\tfrac{1}{3}\beta_{S},\label{eq:ns_2nd}\\
n_{T}&\simeq-2\eps_{1}-s_{T}-s_{c_{T}}-2\eps_{1}^{2}-s_{T}\eps_{1}-\tfrac{1}{3}\beta_{T},\label{eq:nT_2nd}
\end{align}
which, after substitution of~\eqref{eq:eps1_theta}, \eqref{eq:sA_theta}, yields explicit closed-form functions of $\theta$ and of the Gauss--Bonnet parameters $(\delta_{1},\delta_{2})$. In the GR limit $\delta_{1},\delta_{2}\to0$ they reduce to
\begin{equation}
n_{T}\to-2\eps_{1}=-6\sin^{2}\theta,\qquad n_{s}-1\to-2\eps_{1}-\eps_{2},
\label{eq:tilts_GR}
\end{equation}
as expected.

The runnings of the spectral indices read
\begin{equation}
\alpha_{s}\simeq-2\eps_{1}\eps_{2}-\beta_{S}-\beta_{c_{S}},\qquad \alpha_{T}\simeq-2\eps_{1}\eps_{2}-\beta_{T}-\beta_{c_{T}},
\label{eq:alpha_def}
\end{equation}
with $\beta_{c_{A}}=\dot{s}_{c_{A}}/H$.

Moreover, using the parametrization
\begin{equation}
\frac{d}{dN} = \frac{\dot{\theta}}{H} \, \partial_{\theta} + \frac{\dot{H}}{H^{2}} \, \partial_{\ln H},
\end{equation}
combined with Eqs.~\eqref{eq:Hdot_theta} and \eqref{eq:thetadot_theta}, one obtains \(\alpha_{s}\) and \(\alpha_{T}\) as explicit functions of \((H, \theta, V(\phi), \xi(\phi))\) and their derivatives.

\section{Starobinsky inflation with Einstein--Gauss--Bonnet corrections}\label{sec:specific_model}

Starobinsky inflation~\cite{Starobinsky:1980te,DeFelice:2010aj,Martin:2013tda} is one of the oldest and observationally most successful inflationary scenarios. It originated as a purely geometric model in which the gravitational action is supplemented with an $R^{2}$ term, $\mathcal{L}\propto R+R^{2}/(6m^{2})$, and the accelerated expansion is sourced by the additional scalar degree of freedom that this higher-derivative theory propagates. In the Einstein frame the model is equivalent to a canonical scalar field with the plateau potential $V(\phi)=V_{0}\bigl(1-e^{-\sqrt{2/3}\phi}\bigr)^{2}$~\cite{Starobinsky:1980te,DeFelice:2010aj,Martin:2013tda}, which delivers the universal predictions $n_{s}=1-2/N_{*}$ and $r=12/N_{*}^{2}$ at the pivot and provides the $\alpha=1$ representative of the $\alpha$-attractor family~\cite{Kallosh:2013yoa,Kallosh:2013hoa,Kallosh:2014rga,Galante:2014ifa}. The Planck and Planck+BICEP/Keck analyses identified the Starobinsky potential as one of the preferred single-field models~\cite{Planck:2018vyg,BICEP:2021xfz}; recent ACT~DR6 results~\cite{AtacamaCosmologyTelescope:2025nti,AtacamaCosmologyTelescope:2025blo} push the central value of $n_{s}$ upward and sit roughly $1$--$2\sigma$ above the pure Starobinsky prediction, motivating modifications that increase $n_{s}$ while keeping $r\sim10^{-3}$.

Embeddings of the Starobinsky model in Einstein--Gauss--Bonnet gravity have been considered in light of the ACT~DR6 release in Refs.~\cite{Yogesh:2025wak}. These analyses confirm that a non-minimal Gauss--Bonnet coupling provides a controlled mechanism to shift the Starobinsky prediction in the $(n_{s},r)$ plane while preserving the model's plateau structure and its built-in matter-like reheating phase. In what follows we therefore adopt the Starobinsky potential supplemented by the simplest non-trivial Gauss--Bonnet coupling and analyse it within the phase-$\theta$ formalism of Sec.~\ref{sec:theta}.

The coupling function is taken in its minimal non-trivial form $\xi(\phi)=\xi_{0}\phi$. This choice is motivated by three physical arguments. First, $\xi(\phi)=\xi_{0}\phi$ arises naturally in heterotic-string compactifications, where the dilatonic modulus controls the leading higher-curvature correction to the gravitational sector in the weak-coupling limit~\cite{Kanti:1995vq}. Second, the linear coupling contains a single dimensionful parameter $\xi_{0}$ of dimension $\Mpl^{-2}$, which directly measures the strength of the Gauss--Bonnet effect on inflationary observables and so allows a transparent one-parameter scan against the data. Third, the sign choice $\xi_{0}>0$ gives $\delta_{1}=4H\xi_{0}\dot{\phi}<0$ along the inflationary trajectory ($\dot{\phi}<0$), producing a \emph{positive} shift $\Delta n_{s}=-2\delta_{1}>0$ relative to the GR Starobinsky prediction $n_{s}^{\rm GR}=1-2/N_{*}$~\cite{Koh:2014bka,Koh:2016abf}. This is precisely the sign required to move the Starobinsky model into the central part of the ACT~DR6 $1\sigma$ window for $n_{s}$ without inflating $r$. The other types of the coupling function $\xi=\xi(\phi)$ considered, for example, in~\cite{Guo:2010jr,Koh:2014bka,Oikonomou:2023bli} produce similar shifts at the cost of additional parameters and are not required by the data at the present precision.

\subsection{Model setup and closed-form \texorpdfstring{$(\theta,H)$}{}representation}\label{subsec:model_def}

We apply the formalism of Sec.~\ref{sec:theta} to the EGB realization of the Starobinsky inflation
\begin{align}
\label{eq:specific_model}
&V(\phi)=V_{0}\bigl(1-e^{-\sqrt{2/3}\,\phi}\bigr)^{2},\\
\label{XI}
&\xi(\phi)=\xi_{0}\phi,
\end{align}
whose two parameters $(V_{0},\xi_{0})$ are listed together with the initial conditions used by the numerical integration in Table~\ref{tab:initial_conditions}. Two structural features of~\eqref{eq:specific_model} are essential for the analysis that follows: (i) the plateau $V(\phi\gtrsim\Mpl)\to V_{0}$ delivers the universal pair $n_{s}=1-2/N_{*}$, $r=12/N_{*}^{2}$ at leading order and yields naturally small $r$; (ii) the quadratic minimum $V(\phi\to 0)\approx\tfrac{2}{3}V_{0}\phi^{2}=\tfrac{1}{2}m_{\phi}^{2}\phi^{2}$ with $m_{\phi}^{2}=\tfrac{4}{3}V_{0}$ secures the coherent-oscillation reheating relation $\bar{w}_{\rm rh}\to 0^{+}$~\cite{Turner:1983he}.

\label{subsec:model_closed}%
For $\xi=\xi_{0}\phi$ the identity $\dot{\xi}=\xi_{0}\dot{\phi}$ combined with the parametrization~\eqref{eq:theta_param_intro} gives a self-consistent algebraic equation
\begin{equation}
\dot{\xi}=-\sqrt{6}\,\xi_{0}H\sqrt{1-4H\dot{\xi}}\sin\theta.
\end{equation}
Squaring produces a quadratic in $\dot{\xi}$ whose physical root (the one that satisfies the unsquared equation with $\dot{\phi}<0$) is
\begin{equation}
\dot{\xi}(\theta,H)\;=\;-12\,\xi_{0}^{2}H^{3}\sin^{2}\theta\;-\;\sqrt{6}\,\xi_{0}H\sin\theta\,\sqrt{1+24\,\xi_{0}^{2}H^{4}\sin^{2}\theta}.
\label{eq:dxi_starlin}
\end{equation}
Introduce the auxiliaries
\begin{align}
\rho(\theta,H)&\;\equiv\;\xi_{0}H\sin\theta\,\sqrt{1+24\,\xi_{0}^{2}H^{4}\sin^{2}\theta},\label{eq:rho_def}\\
\mathcal{D}_{0}(\theta,H)&\;\equiv\;1-4H\dot{\xi}\;=\;1+48\,\xi_{0}^{2}H^{4}\sin^{2}\theta+4\sqrt{6}\,H\rho.\label{eq:D0_def}
\end{align}
Equations~\eqref{eq:rho_def}--\eqref{eq:D0_def} are exact; $\rho$ is the radical appearing in~\eqref{eq:dxi_starlin} and $\mathcal{D}_{0}=1-\delta_{1}$.

The parametrization~\eqref{eq:theta_param_intro} gives the potential along the trajectory as
\begin{equation}
V(\theta,H)\;=\;3H^{2}\,\mathcal{D}_{0}(\theta,H)\,\cos^{2}\theta.
\label{eq:V_starlin}
\end{equation}
Equating~\eqref{eq:V_starlin} with the Starobinsky form $V=V_{0}(1-e^{-\sqrt{2/3}\phi})^{2}$ and selecting the inflationary plateau branch ($\phi>0$, equivalently $V<V_{0}$) gives the closed-form trajectory
\begin{equation}
\phi(\theta,H)\;=\;-\sqrt{\tfrac{3}{2}}\,\ln\!\Bigl[\,1-\sqrt{V(\theta,H)/V_{0}}\,\Bigr],
\label{eq:phi_starlin}
\end{equation}
and, using $1-e^{-\sqrt{2/3}\phi}=\sqrt{V/V_{0}}$, we obtain
\begin{equation}
V_{,\phi}(\theta,H)\;=\;2\sqrt{\tfrac{2}{3}}\,\sqrt{V(\theta,H)}\,\Bigl[\sqrt{V_{0}}-\sqrt{V(\theta,H)}\Bigr].
\label{eq:Vphi_starlin}
\end{equation}

Note in particular the compact form of the leading factor of $\dot{\theta}$:
\begin{equation}
\frac{V_{,\phi}}{\sqrt{2V}}\;=\;\frac{2}{\sqrt{3}}\Bigl[\sqrt{V_{0}}-\sqrt{V}\Bigr]\;=\;\frac{2\sqrt{V_{0}}}{\sqrt{3}}-2H\cos\theta\,\sqrt{\mathcal{D}_{0}}\,.
\label{eq:Vphi_over_sqrt2V_starlin}
\end{equation}

Three exact algebraic relations close the system. The chain rule on $\dot{\xi}(\theta,H)$ gives
\begin{equation}
\ddot{\xi} \;=\; \frac{\partial\dot{\xi}}{\partial\theta}\,\dot{\theta} + \frac{\partial\dot{\xi}}{\partial H}\,\dot{H};
\label{eq:chain_dxi_starlin}
\end{equation}
the second Friedmann Eq.~\eqref{eq:Hdot_theta} reads
\begin{equation}
\dot{H} \;=\; -3H^{2}\sin^{2}\theta + \frac{2H^{2}(\ddot{\xi}-H\dot{\xi})}{\mathcal{D}_{0}};\label{eq:Hdot_starlin}
\end{equation}
and the phase Eq.~\eqref{eq:thetadot_theta} reads
\begin{equation}
\dot{\theta} = \frac{V_{,\phi}}{\sqrt{2V}} - \frac{2H\ddot{\xi}}{\mathcal{D}_{0}}\cot\theta
-\left[3H^{2}\sin^{2}\theta + \frac{2H^{3}\dot{\xi}-2H^{2}\ddot{\xi}}{\mathcal{D}_{0}}\right]\left[\frac{1}{H} - \frac{2\dot{\xi}}{\mathcal{D}_{0}}\right]\cot\theta.\label{eq:thetadot_starlin}
\end{equation}
The partials $\partial_{\theta}\dot{\xi}$ and $\partial_{H}\dot{\xi}$ are obtained from~\eqref{eq:dxi_starlin}:
\begin{align}
\frac{\partial\dot{\xi}}{\partial\theta} &= -24\,\xi_{0}^{2}H^{3}\sin\theta\cos\theta-\frac{\sqrt{6}\,\xi_{0}H\cos\theta\bigl(1+48\,\xi_{0}^{2}H^{4}\sin^{2}\theta\bigr)}{\sqrt{1+24\,\xi_{0}^{2}H^{4}\sin^{2}\theta}},
\label{eq:dxi_theta_partial}\\
\frac{\partial\dot{\xi}}{\partial H} &= -36\,\xi_{0}^{2}H^{2}\sin^{2}\theta-\frac{\sqrt{6}\,\xi_{0}\sin\theta\bigl(1+72\,\xi_{0}^{2}H^{4}\sin^{2}\theta\bigr)}{\sqrt{1+24\,\xi_{0}^{2}H^{4}\sin^{2}\theta}}.
\label{eq:dxi_H_partial}
\end{align}
System~\eqref{eq:chain_dxi_starlin}--\eqref{eq:thetadot_starlin} is linear in $(\ddot{\xi},\dot{H},\dot{\theta})$ and, after substitution of~\eqref{eq:Vphi_starlin}, depends only on $(\theta,H)$ and the model parameters $(\xi_{0},V_{0})$. Its solution is a triple of explicit algebraic functions
\begin{equation}
\bigl(\ddot{\xi},\,\dot{H},\,\dot{\theta}\bigr) \;=\; \bigl(\ddot{\xi}_{*}(\theta,H),\,\dot{H}_{*}(\theta,H),\,\dot{\theta}_{*}(\theta,H)\bigr),
\label{eq:closure_solutions}
\end{equation}
which are bilinear in $V_{,\phi}$ and the partials~\eqref{eq:dxi_theta_partial}--\eqref{eq:dxi_H_partial}. Direct verification against the Klein--Gordon Eq.~\eqref{eq:EGB_KG} (with $\xi_{,\phi\phi}=0$ for the linear coupling) gives identical results at machine precision.

With $(\dot{\xi},\ddot{\xi},\dot{H},\dot{\theta})$ in closed form, the slow-roll parameters become explicit algebraic functions of $(\theta,H)$:
\begin{align}
&\delta_{1}(\theta,H)\;=\;4H\dot{\xi}\;=\;-(\mathcal{D}_{0}-1),\\
&\delta_{2}(\theta,H)\;=\;\frac{\ddot{\xi}_{*}(\theta,H)}{H\dot{\xi}(\theta,H)},\\
&\Delta(\theta,H)\;=\;\frac{\delta_{1}}{1-\delta_{1}}.
\label{eq:delta_starlin}
\end{align}
The Hubble slow-roll parameter follows either from the identity~\eqref{eq:eps1_theta} or directly from $\eps_{1}=-\dot{H}_{*}/H^{2}$, and these two routes agree identically. The second Hubble parameter is the chain-rule derivative of the first:
\begin{equation}
\eps_{2}(\theta,H)\;=\;\frac{1}{H\eps_{1}}\!\left[\partial_{\theta}\eps_{1}\cdot\dot{\theta}_{*}+\partial_{H}\eps_{1}\cdot\dot{H}_{*}\right]\!.
\label{eq:eps2_starlin}
\end{equation}

The exact $Q_{T}, c_{T}^{2}, Q_{S}, c_{S}^{2}$ of~\eqref{eq:QT_theta}--\eqref{eq:cS_theta} become explicit functions of $(\theta,H)$ after substitution of $\delta_{1}, \delta_{2}, \eps_{1}$. Their slow-roll growth rates $s_{A}\equiv\dot{Q}_{A}/(HQ_{A})$ and $s_{c_{A}}\equiv\dot{c}_{A}/(Hc_{A})$ likewise follow by chain rule, e.g.
\begin{equation}
s_{S}(\theta,H)\;=\;\frac{1}{HQ_{S}}\!\left[\partial_{\theta}Q_{S}\cdot\dot{\theta}_{*}+\partial_{H}Q_{S}\cdot\dot{H}_{*}\right]\!,
\label{eq:sS_starlin}
\end{equation}
and analogously for $s_{T}, s_{c_{T}}, s_{c_{S}}$. The runnings $\beta_{A}=\dot{s}_{A}/H$ are obtained by one further chain rule on $s_{A}(\theta,H)$. With these in hand the Hankel indices~\eqref{eq:nu_indices} are evaluated at the pivot as exact functions of $(\theta_{*},H_{*},\xi_{0},V_{0})$, and the dimensionless spectra~\eqref{eq:PS_full}--\eqref{eq:PT_full} together with the tilts and runnings~\eqref{eq:tilts_GR}--\eqref{eq:alpha_def} follow without any further approximation.

The end-of-inflation locus is set by $\eps_{1}(\theta_{e})=1$. From the identity~\eqref{eq:eps1_theta}, $\sin^{2}\theta_{e}=\tfrac{1}{3}[1-\tfrac{1}{2}\Delta(1-\delta_{2})]_{e}$, which reduces to $\sin^{2}\theta_{e}=1/3$ in the GR limit ($\theta_{e}^{\rm GR}=\arcsin(1/\sqrt{3})\simeq 35.26^{\circ}$). The Gauss--Bonnet correction to $\theta_{e}$ is $\mathcal{O}(\Delta)$ and, for the specific model studied here, evaluates to $\Delta\theta_{e}\simeq +3\times 10^{-4}~\mathrm{rad}$; the numerical value $\theta_{e}=0.6158~\mathrm{rad}\simeq 35.28^{\circ}$ is recorded in Table~\ref{tab:initial_conditions} together with $|\delta_{1}|_{e}\simeq 3.6\times 10^{-3}$ and $|\delta_{2}|_{e}\simeq 5.4\times 10^{-1}$. The phase $\theta$ continues to evolve past $\theta_{e}$ during the post-inflationary oscillations and reaches $\theta=\pi/2$ transiently at the crossings of the potential minimum ($V=0$, cf.\ Sec.~\ref{subsec:obs_theta}).

 \begin{figure}[H]
\centering
\includegraphics[width=0.99\textwidth]{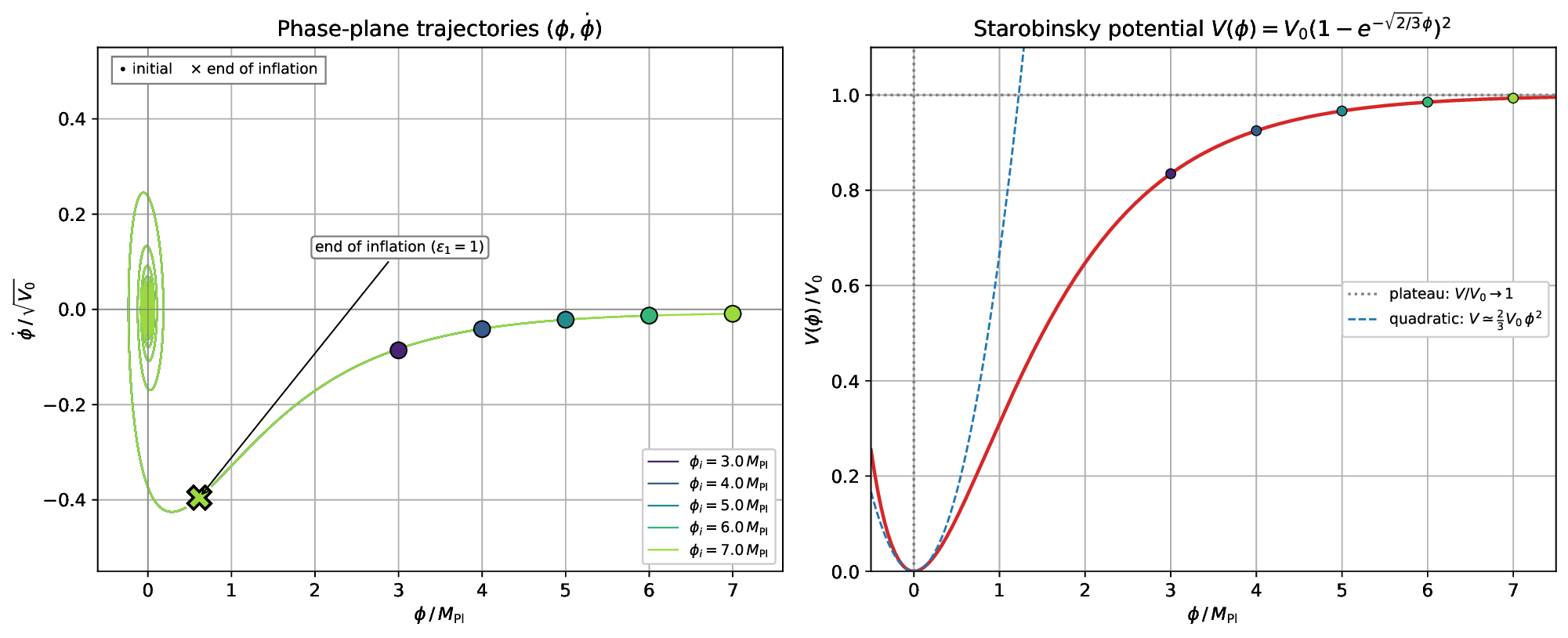}
\caption{Phase-space trajectories of the inflaton for the Sratobinsky inflation. {\bf Left Panel:} $(\phi,\dot{\phi})$ phase-plane trajectories for plateau initial conditions $\phi_{i}\in\{3,4,5,6,7\}\,\Mpl$, with circles marking the initial points and ``$\times$'' the end-of-inflation locus $\eps_{1}=1$. {\bf Right Panel:} Starobinsky potential $V(\phi)/V_{0}$ (red solid) vs.\ the quadratic-minimum approximation $V\simeq\tfrac{2}{3}V_{0}\phi^{2}$ (blue dashed), with the same colour-coded initial conditions. $\dot{\phi}$ is shown in units of $\sqrt{V_{0}}$.}
\label{fig:phase_space}
\end{figure}

In the conventional $(\phi,\dot{\phi})$ phase plane the trajectories of single-field inflaton models are organized around the structure of $V(\phi)$ and admit a coarse dynamical-systems classification. The two relations of~\eqref{eq:theta_param_intro} furnish, in particular, an explicit injective map from the EGB phase plane to the half-strip $\theta\in[0,\pi/2]$ along the inflationary trajectory and to a full circle $\theta\in[0,2\pi)$ during the oscillatory phase. The trajectory is of the standard inflationary-attractor type: the inflaton starts on the de~Sitter-like inflationary plateau ($\dot{\phi}\to 0^{-}$, $\phi\gg \Mpl$, $\theta\to 0$), is monotonically drawn towards the quadratic minimum of the Starobinsky potential (\ref{eq:specific_model}), crosses the end-of-inflation locus $\eps_{1}=1$ at $\theta=\theta_{e}$ with
\begin{equation}
\sin^{2}\theta_{e}=\tfrac{1}{3}[1-\tfrac{1}{2}\Delta(1-\delta_{2})]_{\theta_{e},H_{e}}
\end{equation}
from~\eqref{eq:eps1_theta} (which reduces to $\sin^{2}\theta_{e}=1/3$, $\theta_{e}\simeq 35.26^{\circ}$ in the GR limit and receives an $\mathcal{O}(\Delta)$ Gauss--Bonnet correction otherwise), and develops damped oscillations around $\phi=0$ that asymptote to an attracting focus at the origin of the $(\phi,\dot{\phi})$ plane. The EGB-coupling deforms the orbit through the prefactor $\sqrt{1-4H\dot{\xi}}$ in~\eqref{eq:theta_param_intro} without changing its topological type. Fig.~\ref{fig:phase_space} shows the $(\phi,\dot{\phi})$ phase-plane trajectories obtained from the full background integration for five different plateau initial conditions $\phi_{i}\in[3,7]\,\Mpl$: all orbits fall onto a common attractor branch on the inflationary plateau (the slow-roll attractor) and then spiral inward to the origin during the post-inflationary oscillations, irrespective of the specific starting point, confirming the attractor-type dynamical classification claimed above. The Starobinsky potential $V(\phi)$ is shown alongside for orientation, together with the quadratic minimum approximation $V\simeq\tfrac{2}{3}V_{0}\phi^{2}$ relevant to the reheating regime.

\subsection{Numerical methodology}\label{subsec:model_method}

The background equations~\eqref{eq:EGB_F1}--\eqref{eq:EGB_KG} are integrated in full using the explicit Dormand--Prince 8(7) Runge--Kutta scheme~\cite{dormand1980family,prince1981high} with adaptive step, relative tolerance $10^{-11}$ and absolute tolerance $10^{-12}$. The end of inflation is fixed by $\eps_{1}(t_{e})=1$.
The maximum relative residual of the Hamiltonian constraint, evaluated over the entire inflationary trajectory
\begin{equation}\label{eq:hamiltonian_residual}
\frac{\bigl|3H^{2} - \bigl(\tfrac{1}{2}\dot{\phi}^{2} + V + 12H^{3}\dot{\xi}\bigr)\bigr|}{3H^{2}}\leq 10^{-10},
\end{equation}
which confirms the numerical accuracy of the solution.

The integration is initialized deep on the inflationary plateau at $\phi_{i} = 6.0\,M_{\mathrm{pl}}$, taking into account dynamic equation
\begin{equation}\label{ad7}
3H^{2}\bigl(1 - 4H\xi_{,\phi}\dot{\phi}\bigr) = \frac{1}{2}\dot{\phi}^{2} + V,
\end{equation}
is solved jointly with the EGB slow-roll attractor
\begin{equation}\label{ad8}
\dot{\phi} \simeq -\frac{V_{,\phi} + 12H^{4}\xi_{,\phi}}{3H},
\end{equation}
at \(\phi = \phi_{i}\) to a relative tolerance of \(10^{-12}\).

At this point on the plateau the two driving terms of the attractor relation are comparable, $V_{,\phi}(\phi_{i})\simeq 2.22\times 10^{-12}\,\Mpl^{3}$ and $12H_{i}^{4}\xi_{,\phi}(\phi_{i})\simeq 1.92\times 10^{-12}\,\Mpl^{3}$, so neglecting the Gauss--Bonnet contribution at the initialization would distort $H_{i}$ at the percent level. The converged values of $H_{i}$, $\dot{\phi}_{i}$ and the corresponding initial phase $\theta_{i}$ inferred from the polar parametrization~\eqref{eq:theta_param_intro} are listed in Table~\ref{tab:initial_conditions} in both reduced-Planck and physical units; the GB-modified Friedmann prefactor $(1-\delta_{1})_{i}\simeq 1+2.4\times 10^{-4}$ at initialization documents the size of the Gauss--Bonnet contribution at the start of the integration. The choice $\theta_{i}\to 0$ as the natural start of the trajectory follows the convention of the GR phase-$\theta$ parametrization of~\cite{Kaur:2023wos}, where $\theta=0$ identifies the deep slow-roll plateau.

\begin{table}[h]
\centering
\caption{Initial conditions used for the background integration of Eqs.~\eqref{eq:EGB_F1}--\eqref{eq:EGB_KG} and for the subsequent Mukhanov--Sasaki integration~\eqref{eq:MS}. Construction procedure: see Sec.~\ref{subsec:model_method}.}\label{tab:initial_conditions}
\begin{tabular}{@{}lcc@{}}
\toprule
Quantity & Reduced Planck units & Physical units\\
\midrule
$\phi_{i}$ & $6.0\,\Mpl$ & $1.46\times 10^{19}~\mathrm{GeV}$\\
$\dot{\phi}_{i}$ & $-1.775\times 10^{-7}\,\Mpl^{2}$ & $-1.052\times 10^{30}~\mathrm{GeV}^{2}$\\
$H_{i}$ & $7.771\times 10^{-6}\,\Mpl$ & $1.892\times 10^{13}~\mathrm{GeV}$\\
$V(\phi_{i})$ & $1.812\times 10^{-10}\,\Mpl^{4}$ & $6.367\times 10^{63}~\mathrm{GeV}^{4}$\\
$\theta_{i}$ & $9.32\times 10^{-3}\;\mathrm{rad}\;(\to 0)$ & $0.534^{\circ}$\\
$\delta_{1}(t_{i})$ & $-2.41\times 10^{-4}$ & --\\
$V_{0}$ & $1.83924\times 10^{-10}\,\Mpl^{4}$ & $6.466\times 10^{63}~\mathrm{GeV}^{4}\;\bigl[V_{0}^{1/4}=8.967\times 10^{15}~\mathrm{GeV}\bigr]$\\
$\xi_{0}$ & $4.3768\times 10^{7}\,\Mpl^{-2}$ & $7.382\times 10^{-30}~\mathrm{GeV}^{-2}$\\
\bottomrule
\end{tabular}
\end{table}

The scalar and tensor power spectra are computed by direct numerical integration of the Mukhanov--Sasaki Eq.~\eqref{eq:MS} in conformal time. The exact coefficients $c_{T}^{2}(\eta),c_{S}^{2}(\eta),z_{T}(\eta),z_{S}(\eta)$ of Refs.~\cite{Hwang:2005hb,Koh:2014bka,Koh:2016abf,Koh:2018qcy} are used. Bunch--Davies initial conditions~\cite{Baumann:2014nda} are set at $c_{A}k|\eta|=500$ and the integration is continued to $c_{A}k|\eta|=10^{-3}$, deep on super-horizon scales, on a grid of 41 modes in $k/k_{*}\in[e^{-2},e^{2}]$. The spectral indices and runnings are extracted from a quadratic regression of $\ln\Delta_{A}^{2}(k)$ in $\ln k$. The amplitude is normalized to the value $\Delta_{s}^{2}(k_{*})=A_{s}=2.1\times10^{-9}$~\cite{AtacamaCosmologyTelescope:2025nti,Planck:2018vyg} by iteratively rescaling $V_{0}$.

\subsection{Background trajectory and physical viability}\label{subsec:model_bg}

\begin{figure}[H]
\centering
\includegraphics[width=0.73\textwidth]{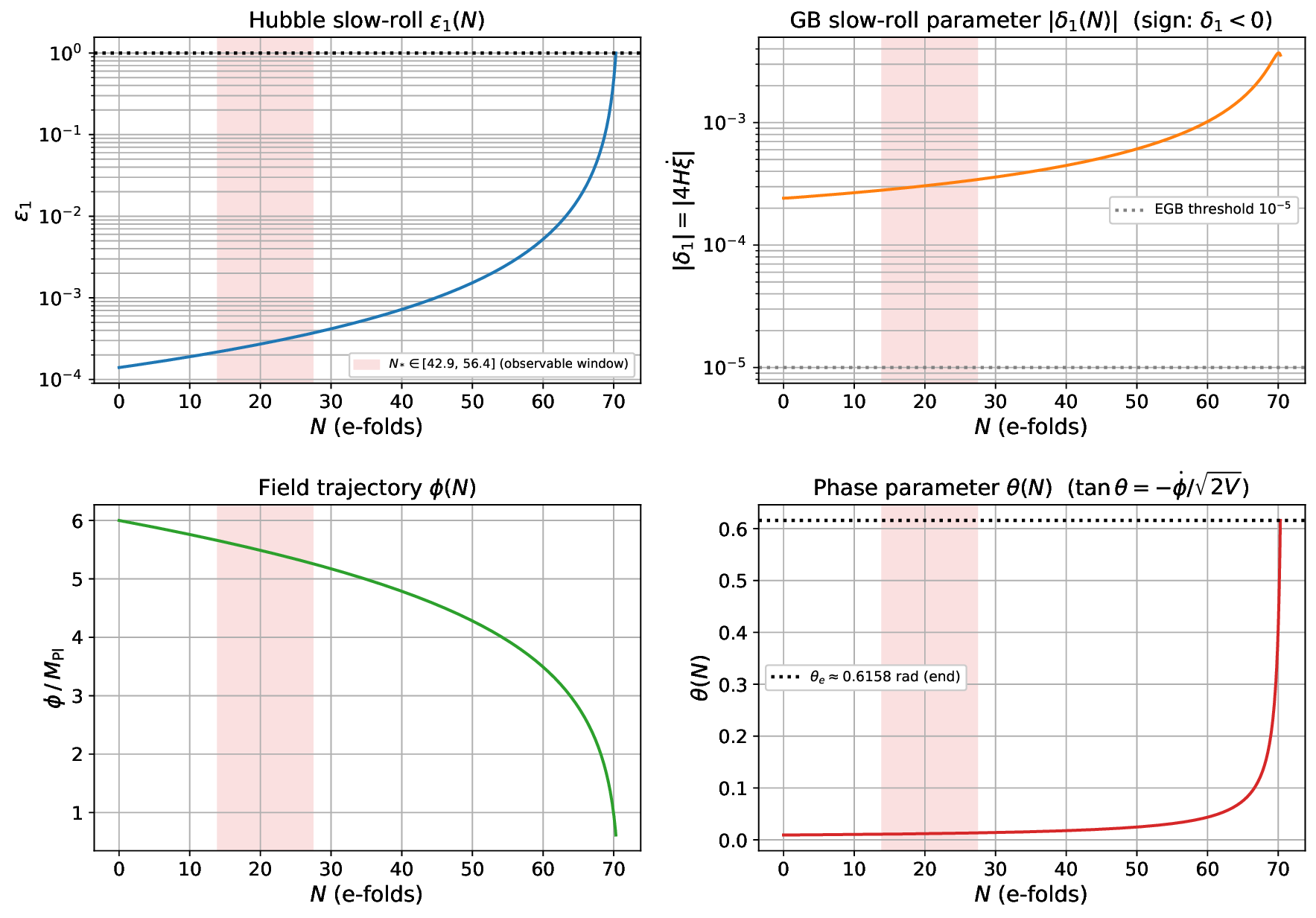}
\caption{Background evolution of the model~\eqref{eq:specific_model}. Top left: $\eps_{1}(N)$ (blue); the gray dotted line marks $\eps_{1}=1$. Top right: $|\delta_{1}(N)|$ (orange); gray dotted line at $|\delta_{1}|=10^{-5}$ (EGB/GR boundary). Bottom left: $\phi(N)$ (green). Bottom right: $\theta(N)$ (red), with $\theta_{e}\simeq 0.6158~\mathrm{rad}$ as gray dotted line. The pink band marks the pivot window $N_{*}\in[42.90,\,56.35]$ (cf.\ Sec.~\ref{subsec:model_Nstar}). End of inflation: $N_{e}=70.30$.}
\label{fig:model_bg}
\end{figure}

Figure~\ref{fig:model_bg} shows the evolution of the slow-roll parameters, the inflaton $\phi(N)$ and the phase $\theta(N)$ throughout inflation; the end of inflation is reached at $N_{e}=70.30$ e-folds. At the pivot $N_{*}=55$ one has $\eps_{1}=2.28\times10^{-4}$, $|\delta_{1}|=2.86\times10^{-4}$, $|\delta_{2}|=1.33\times10^{-2}$ (MS pipeline; the closed-form pipeline gives $|\delta_{2}|=1.24\times10^{-2}$, see Table~\ref{tab:closed_vs_MS}), all well below unity, so the slow-roll expansion is fully under control.

Three additional physical-viability conditions are explicitly checked.
\begin{enumerate}[label=(\arabic*)]
\item Absence of a tensor ghost: $Q_{T}=1-\delta_{1}=1.00029>0$.
\item Absence of a scalar ghost: $Q_{S}(N_{*})=7.37\times10^{-4}>0$.
\item Absence of gradient instabilities: $c_{T}(N_{*})=0.99986$ and $c_{S}(N_{*})=1.0000$. The relevant inflationary-epoch constraint comes from CMB B-mode polarization, $c_{T}<\sqrt{2.85}\simeq 1.69$ at 95\% CL~\cite{Raveri:2014eea} combined with $c_{T}>0.22$ at 95\% CL~\cite{Giare:2020vss}, which the model satisfies by roughly four orders of magnitude ($|c_{T}-1|_{*}\simeq 1.4\times 10^{-4}$). The much tighter GW170817 bound $|c_{T}^{(0)}-1|\lesssim 10^{-15}$ probes the propagation speed at LIGO frequencies and constrains only the late-universe value~\cite{LIGOScientific:2017vwq}; for the linear coupling $\xi(\phi)=\xi_{0}\phi$ studied here, the Gauss--Bonnet contribution to $c_{T}^{2}-1$ is proportional to $\delta_{1}(t)=4H\dot{\xi}$ and is suppressed by $\dot{H}/H^{2}$ at late times, so that at matter--radiation equality $|\delta_{1}|\lesssim 10^{-30}$, comfortably inside the GW170817 bound.
\end{enumerate}

\begin{figure}[H]
\centering
\includegraphics[width=0.73\textwidth]{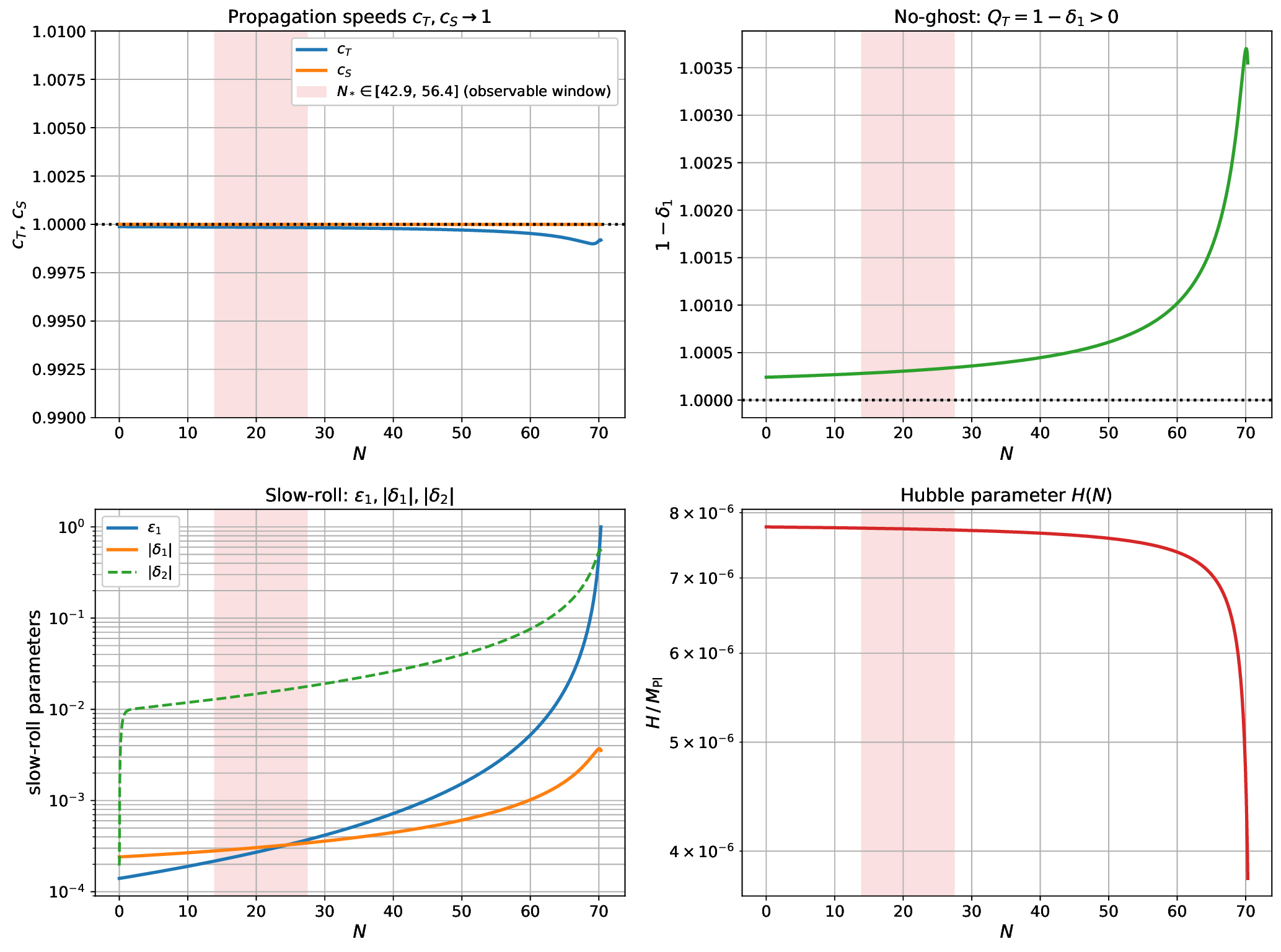}
\caption{Physical-viability checks of the studied model. Top left: propagation speeds $c_{T}(N)$ (blue), $c_{S}(N)$ (orange); both $\equiv 1$ at plot precision. Top right: $Q_{T}(N)=1-\delta_{1}(N)$ (green), positive throughout. Bottom left: slow-roll parameters $\eps_{1}(N)$ (blue), $|\delta_{1}(N)|$ (orange), $|\delta_{2}(N)|$ (green dashed) on log scale. Bottom right: $H(N)$ (red). Pink band: pivot window $N_{*}\in[42.90,\,56.35]$.}
\label{fig:model_stab}
\end{figure}

Their evolution along the trajectory is displayed in Fig.~\ref{fig:model_stab}.

\begin{figure}[H]
\centering
\includegraphics[width=0.95\textwidth]{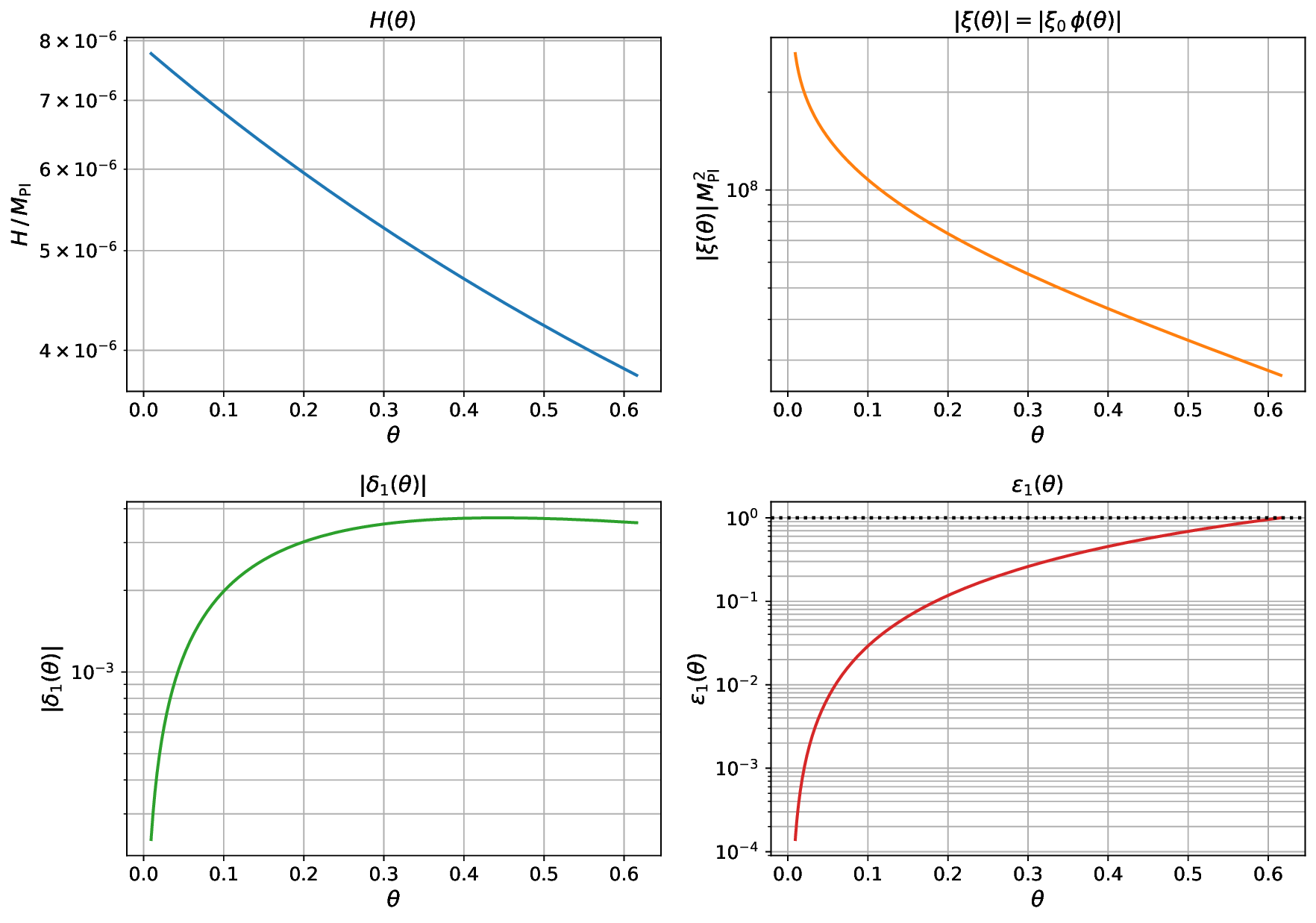}
\caption{Trajectories of the studied model in the phase variable $\theta\in[\theta_{i},\theta_{e}]$. Top left: $H(\theta)$ (blue). Top right: $|\xi(\theta)|=\xi_{0}|\phi(\theta)|$ (orange). Bottom left: $|\delta_{1}(\theta)|$ (green). Bottom right: $\eps_{1}(\theta)$ (red). End-of-inflation value $\theta_{e}\simeq 0.6158~\mathrm{rad}$ taken from Sec.~\ref{subsec:model_def}.}
\label{fig:model_theta}
\end{figure}

Figure~\ref{fig:model_theta} shows the trajectories in $\theta$-coordinates for the Hubble parameter, the coupling $|\xi(\theta)|$, the Gauss--Bonnet parameter $|\delta_{1}(\theta)|$ and the slow-roll parameter $\eps_{1}(\theta)$. The phase $\theta$ evolves monotonically from $\theta\to0$ at the beginning of inflation to the end-of-inflation value $\theta_{e}\simeq 0.6158$, and the Gauss--Bonnet contribution grows toward the end of inflation, naturally driving the breakdown of slow roll.

The numerical Hubble parameter at the pivot is $H_{*}=7.75\times10^{-6}\,\Mpl=1.89\times10^{13}~\mathrm{GeV}$, and at the end of inflation $H_{\rm end}=3.78\times10^{-6}\,\Mpl=9.21\times10^{12}~\mathrm{GeV}$ in physical units, with the reduced Planck mass $\Mpl=(8\pi G)^{-1/2}=2.435\times10^{18}~\mathrm{GeV}$.

\subsection{Allowed window of \texorpdfstring{$N_{*}$}{} and reheating}\label{subsec:model_Nstar}

Two e-fold quantities are used in the rest of this section, with distinct meanings and different reference times. The total number of e-folds $N_{e}\equiv\ln[a(t_{e})/a(t_{i})]$ is the total inflationary expansion accumulated from the initial time $t_{i}$ on the plateau to the end of inflation $t_{e}$ (defined by $\eps_{1}(t_{e})=1$), and for the present model yields $N_{e}=70.30$, comfortably above the canonical minimum $N_{e}\gtrsim 60$. The pivot e-fold $N_{*}\equiv\ln[a(t_{e})/a(t_{*})]$ is instead measured backward from the end of inflation, with $t_{*}$ the time at which the comoving pivot mode $k_{*}=0.05~\mathrm{Mpc}^{-1}$ exits the comoving Hubble radius. We adopt $N_{*}=55$ as the nominal pivot value; the physical reheating window discussed below selects $N_{*}\in[42.90,\,56.35]$, and the gap $N_{i\to *}=N_{e}-N_{*}=15.30$ e-folds of pre-pivot plateau evolution sits comfortably outside the CMB-observable window.

The pivot mode is matched to a number of e-folds $N_{*}$ through the standard relation~\cite{Liddle:2003as,Martin:2010kz}
\begin{equation}
N_{*}\approx 67-\ln\frac{k_{*}}{a_{0}H_{0}}+\tfrac{1}{4}\ln\!\frac{V_{*}^{2}}{\Mpl^{4}\rho_{\rm end}}+\frac{1-3\bar{w}_{\rm rh}}{12(1+\bar{w}_{\rm rh})}\ln\!\frac{\rho_{\rm rh}}{\rho_{\rm end}}.
\label{eq:N_star}
\end{equation}

\begin{figure}[H]
\centering
\includegraphics[width=0.99\textwidth]{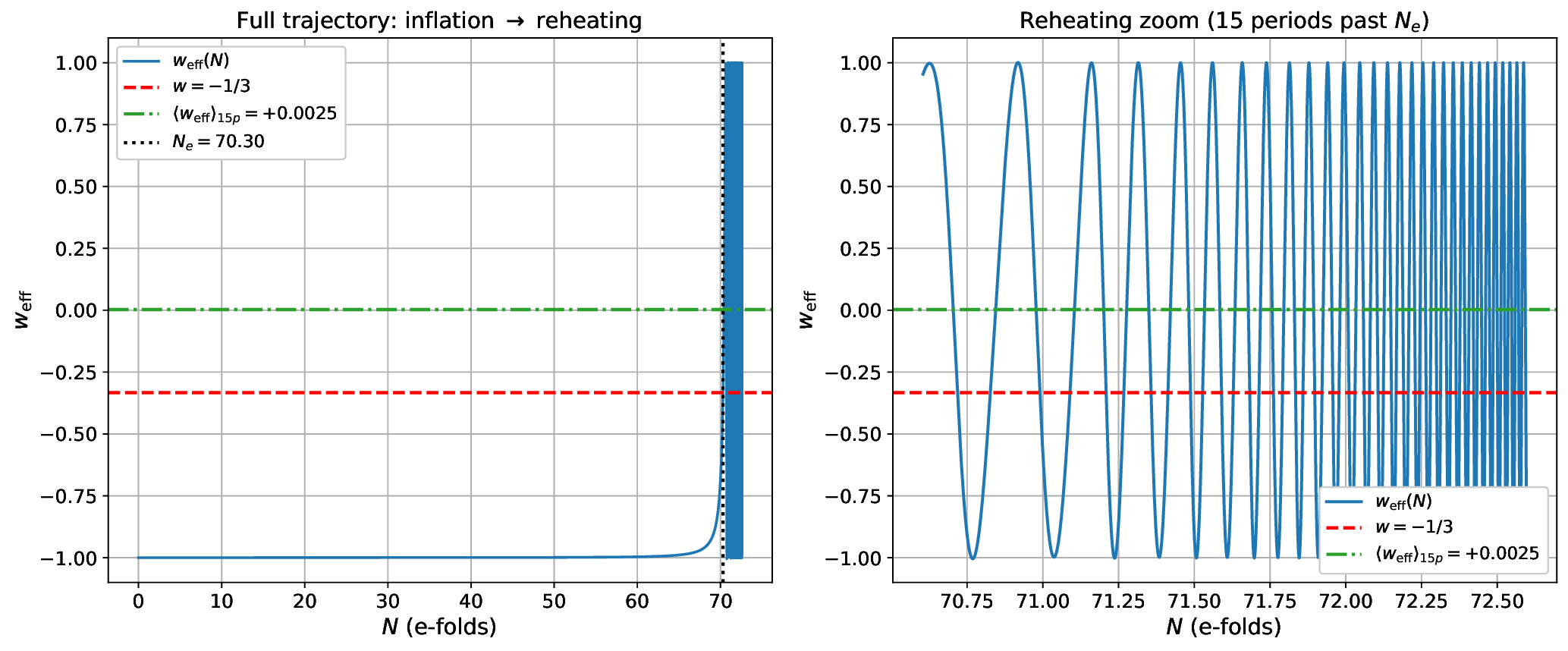}
\caption{Effective equation-of-state parameter $w_{\rm eff}(N)$ of the studied model, computed from Eq.~\eqref{eq:w_eff_theta} along the numerical background trajectory. Left: full trajectory through $15\,T_{\rm osc}$ past $N_{e}$. Right: zoom on the first $\sim 2$ e-folds past $N_{e}$. Red dashed: $w_{\rm eff}=-1/3$. Green dash-dotted: $\langle w_{\rm eff}\rangle_{15p}=+0.0025$ of~\eqref{eq:running_avg}.}
\label{fig:w_eff_window}
\end{figure}

\begin{figure}[H]
\centering
\includegraphics[width=0.78\textwidth]{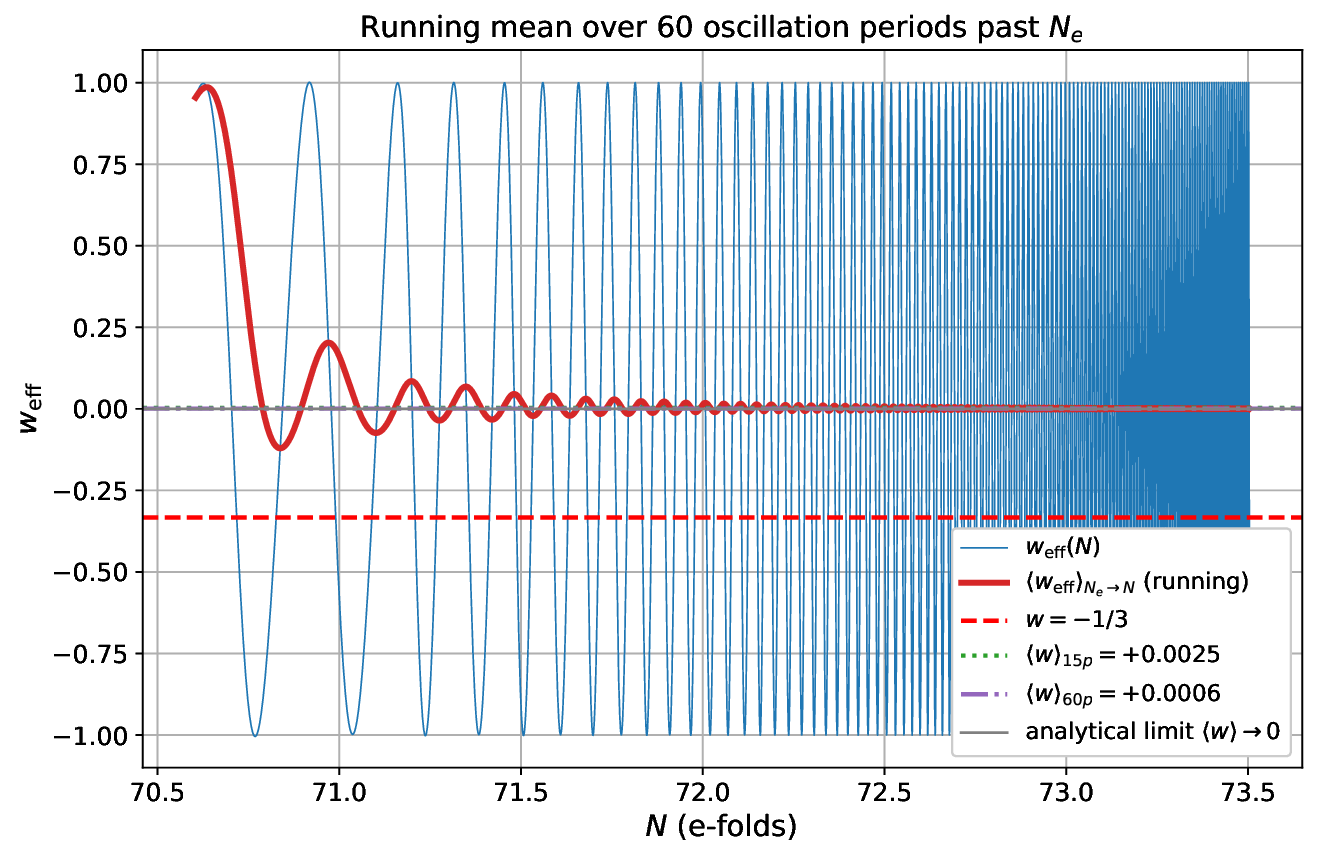}
\caption{Running average $\langle w_{\rm eff}\rangle_{N_{e}\to N}$ of Eq.~\eqref{eq:running_avg} over $60$ inflaton oscillation periods past $N_{e}$ (red solid, on faint blue oscillation envelope). Reference lines: $\langle w_{\rm eff}\rangle_{15p}=+0.0025$ (green dotted), $\langle w_{\rm eff}\rangle_{60p}=+6\times 10^{-4}$ (purple dash-dotted), analytical asymptote $\langle w\rangle\to 0$ (gray solid), $w_{\rm eff}=-1/3$ (red dashed).}
\label{fig:w_eff_running}
\end{figure}

To compare the full numerical integration of Sec.~\ref{subsec:model_method} with the analytical statement~\eqref{eq:wbar_bound}, it is useful to introduce the \emph{running} time average of \(w_{\rm eff}\) accumulated from the end of inflation up to a generic later time \(t(N)\):
\begin{equation}
\langle w_{\rm eff}\rangle_{N_{e}\to N}
\equiv\frac{1}{t(N)-t_{e}}\!\int_{t_{e}}^{t(N)}\!w_{\rm eff}(t')\,dt'.
\label{eq:running_avg}
\end{equation}
This quantity is well-defined for any \(N>N_{e}\) and depends on the chosen upper limit. As the upper limit is pushed further past \(N_{e}\), the running average~\eqref{eq:running_avg} continues to drift toward the analytical asymptote \(\bar w_{\rm eff}\to 0\) of Eq.~\eqref{eq:wbar_bound}, as expected once the oscillation amplitude \(A(t)\) damps.
The numerical values obtained for the model under study are window-dependent: a short window of approximately \(15\) inflaton oscillation periods past \(N_{e}\) yields \(\langle w_{\rm eff}\rangle_{15p}=+0.0025\). A longer window of approximately \(60\) periods gives \(\langle w_{\rm eff}\rangle_{60p}=+6\times 10^{-4}\). These values sit comfortably above the lower bound \(-1/3\) and well below \(+1/3\); the qualitative reheating phenomenology (a matter-like, mildly accelerated approach to the radiation era) is therefore window-independent. Fig.~\ref{fig:w_eff_running} displays the running average~\eqref{eq:running_avg} together with the short- and long-window numerical values and the analytical limit~\eqref{eq:wbar_bound}.
This value, plotted in Figs.~\ref{fig:w_eff_window} and~\ref{fig:w_eff_running}, is used throughout the $N_{*}$ matching. Scanning the reheating temperature from the BBN bound $T_{\rm rh}\gtrsim5~\mathrm{MeV}$~\cite{deSalas:2015glj} to the instantaneous-reheating value $T_{\rm rh}^{\rm inst}\approx2.56\times10^{15}~\mathrm{GeV}$ yields the windows in Table~\ref{tab:Nstar_range}. The physical column $\bar{w}_{\rm rh}=0.0025$ gives the window $N_{*}\in[42.90,\,56.35]$.

\begin{table}[H]
\centering
\caption{Allowed window for $N_{*}$ as a function of reheating temperature $T_{\rm rh}$ and equation of state $\bar{w}_{\rm rh}$. The physical column for the studied model is $\bar{w}_{\rm rh}=0.0025$ (Eq.~\eqref{eq:running_avg}); $\bar{w}_{\rm rh}=1/3$ is shown for reference.}\label{tab:Nstar_range}
\begin{tabular}{@{}lccc@{}}
\toprule
$T_{\rm rh}$ & $N_{*}(\bar{w}_{\rm rh}=0.0025)$ & $N_{*}(\bar{w}_{\rm rh}=1/3)$ & $N_{*}(\bar{w}_{\rm rh}=1)$\\
\midrule
$5~\mathrm{MeV}$ & 42.90 & 56.35 & 69.95\\
$10^{8}~\mathrm{GeV}$ & 50.73 & 56.35 & 62.04\\
$10^{12}~\mathrm{GeV}$ & 53.77 & 56.35 & 58.97\\
$T_{\rm rh}^{\rm inst}\approx2.56\times10^{15}~\mathrm{GeV}$ & 56.35 & 56.35 & 56.35\\
\bottomrule
\end{tabular}
\end{table}

The pivot $N_{*}=55$ lies within this window and corresponds to $T_{\rm rh}\approx7.7\times10^{13}~\mathrm{GeV}$. An additional upper bound on the GB-active phase of inflation follows from the positivity of the Gauss--Bonnet contribution to the energy density, $N_{\rm GB}\leq 2 N_{E}$ with $N_{E}=60$ the canonical observable e-fold count~\cite{Fomin:2020hfh}; numerically $N_{\rm GB}<120$. In the present model $\rho_{\rm GB}\propto H^{3}\dot\xi$ keeps a definite sign throughout the slow-roll trajectory (positive for $\dot\phi<0$, $\xi_{0}>0$), so the running e-fold count defined in Ref.~\cite{Fomin:2020hfh} from $\dot N_{\rm GB}=H\,\Theta(\rho_{\rm GB})$ (with $\Theta$ the Heaviside step) coincides with the geometric $N_{e}=70.30$ to the precision relevant here, and the bound $N_{\rm GB}\leq 2 N_{E}$ is comfortably satisfied.

\subsection{CMB observables and confrontation with ACT~DR6 + BICEP/Keck}\label{subsec:model_cmb}

Direct integration of the Mukhanov--Sasaki equation yields the spectra and their tilts; the resulting observables at the pivot $N_{*}=55$ are summarized in Table~\ref{tab:model_obs}.

\begin{table}[H]
\centering
\renewcommand{\arraystretch}{1.2}
\caption{Observable predictions of the studied model at $N_{*}=55$ from full integration of the Mukhanov--Sasaki equation, confronted with ACT~DR6 + BICEP/Keck~\cite{AtacamaCosmologyTelescope:2025nti,AtacamaCosmologyTelescope:2025blo,DESI:2024mwx,Planck:2018vyg,BICEP:2021xfz}.}\label{tab:model_obs}
\begin{tabular}{@{}lccc@{}}
\toprule
Parameter & Value & Constraint & Status\\
\midrule
$n_{s}$ & $0.9730$ & $0.974\pm0.003$ (ACT~DR6 v2, $1\sigma$) & inside $1\sigma$\\
$r$ & $5.78\times10^{-3}$ & $r<0.038$ (BICEP/Keck, 95\%) & satisfied\\
$\alpha_{s}$ & $-6.06\times10^{-4}$ & $+0.0062\pm0.0052$ (ACT~DR6) & inside $1.4\sigma$\\
$\Delta_{s}^{2}(k_{*})$ & $2.100\times10^{-9}$ & $2.10\times10^{-9}$ (Planck) & satisfied\\
$N_{e}$ & $70.30$ & $N_{\rm GB}<120$~\cite{Fomin:2020hfh} & inside\\
$|\delta_{1}|_{*}$ & $2.86\times10^{-4}$ & $>10^{-5}$ & EGB non-trivial\\
$c_{T}$ & $0.99986$ & \makecell{$c_{T}^{2}<2.85$ (Planck+BICEP2 BB, $95\%$~CL)~\cite{Raveri:2014eea}\\ $c_{T}>0.22$ ($95\%$~CL)~\cite{Giare:2020vss}} & satisfied\\
\bottomrule
\end{tabular}
\renewcommand{\arraystretch}{1.0}
\end{table}

\begin{figure}[H]
\centering
\includegraphics[width=0.85\textwidth]{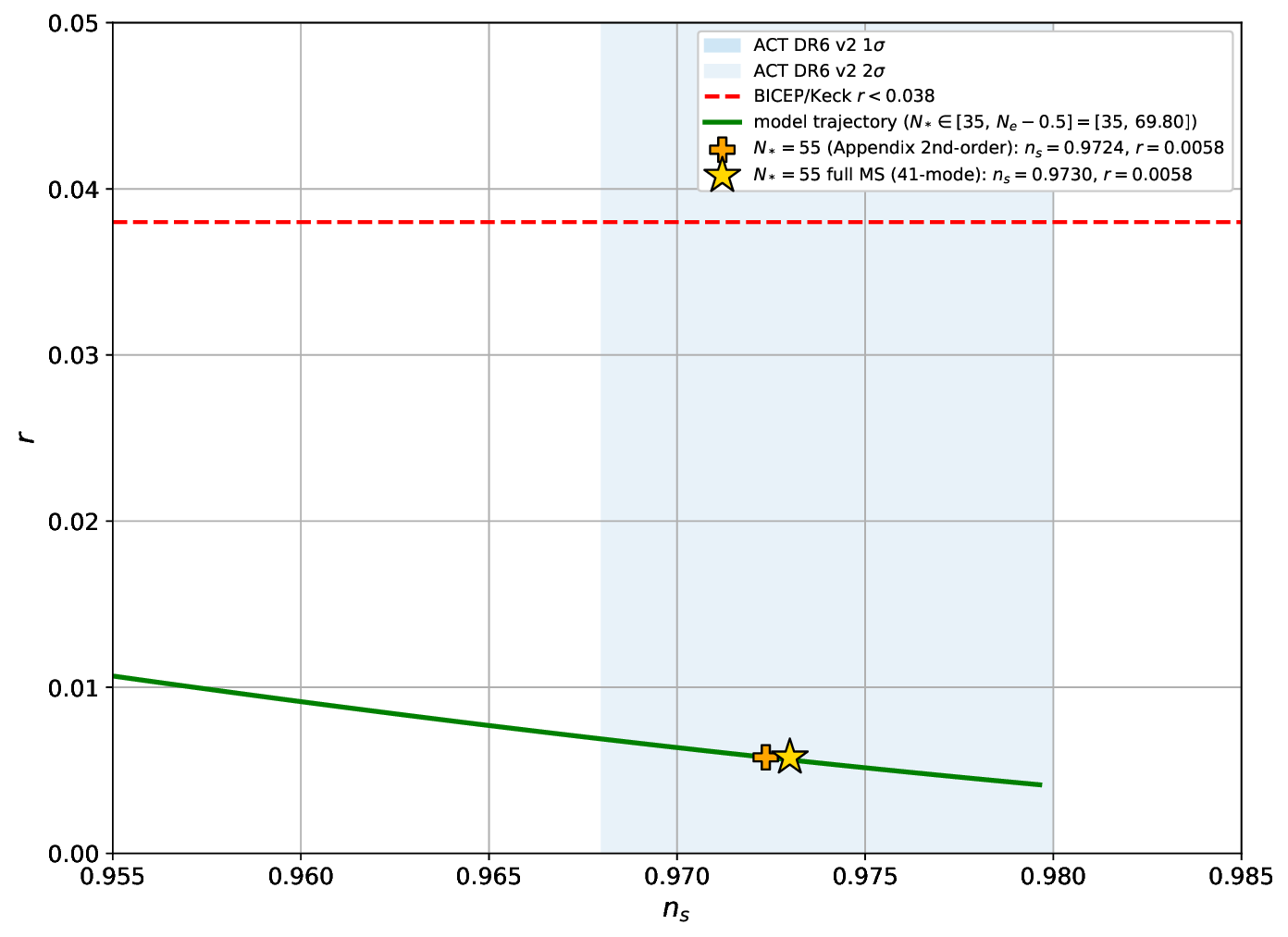}
\caption{Position of the studied model in the $(n_{s},r)$ plane against ACT~DR6 + BICEP/Keck. Green solid line: model trajectory under variation of $N_{*}$; yellow star: pivot $N_{*}=55$ from full MS integration; orange ``+'': pivot value from the second-order analytical formulae. Blue bands: ACT~DR6 v2 $1\sigma$/$2\sigma$ on $n_{s}$; red dashed: BICEP/Keck $r<0.038$.}
\label{fig:model_nsr}
\end{figure}

All observational bounds are simultaneously satisfied, and the Gauss--Bonnet sector is dynamically active: $|\delta_{1}|_{*}$ exceeds the threshold $10^{-5}$ separating the EGB regime from the effective GR limit by more than an order of magnitude. The position of the model in the $(n_{s},r)$ plane is shown in Fig.~\ref{fig:model_nsr}: the green solid curve is the trajectory of the model under variation of $N_{*}\in[35,\,N_{e}-0.5]=[35,\,69.80]$; the yellow star marks the pivot $N_{*}=55$ from the full Mukhanov--Sasaki integration and the orange ``$+$'' marks the same point computed analytically from the second-order phase-$\theta$ formulae of Appendix~\ref{app:derivation}.

For $N_{*}\gtrsim 51$ the trajectory enters the ACT~DR6 $1\sigma$ band on $n_{s}$ and remains inside it throughout the physical reheating window of Table~\ref{tab:Nstar_range} (up to $N_{*}=56.35$), while $r$ stays well below the BICEP/Keck bound across the same range. The pivot $N_{*}=55$ corresponds to $n_{s}=0.9730$, in the central part of the $1\sigma$ region.

\subsection{Comparison of computational methods}\label{subsec:model_methods}

\begin{figure}[H]
\centering
\includegraphics[width=0.99\textwidth]{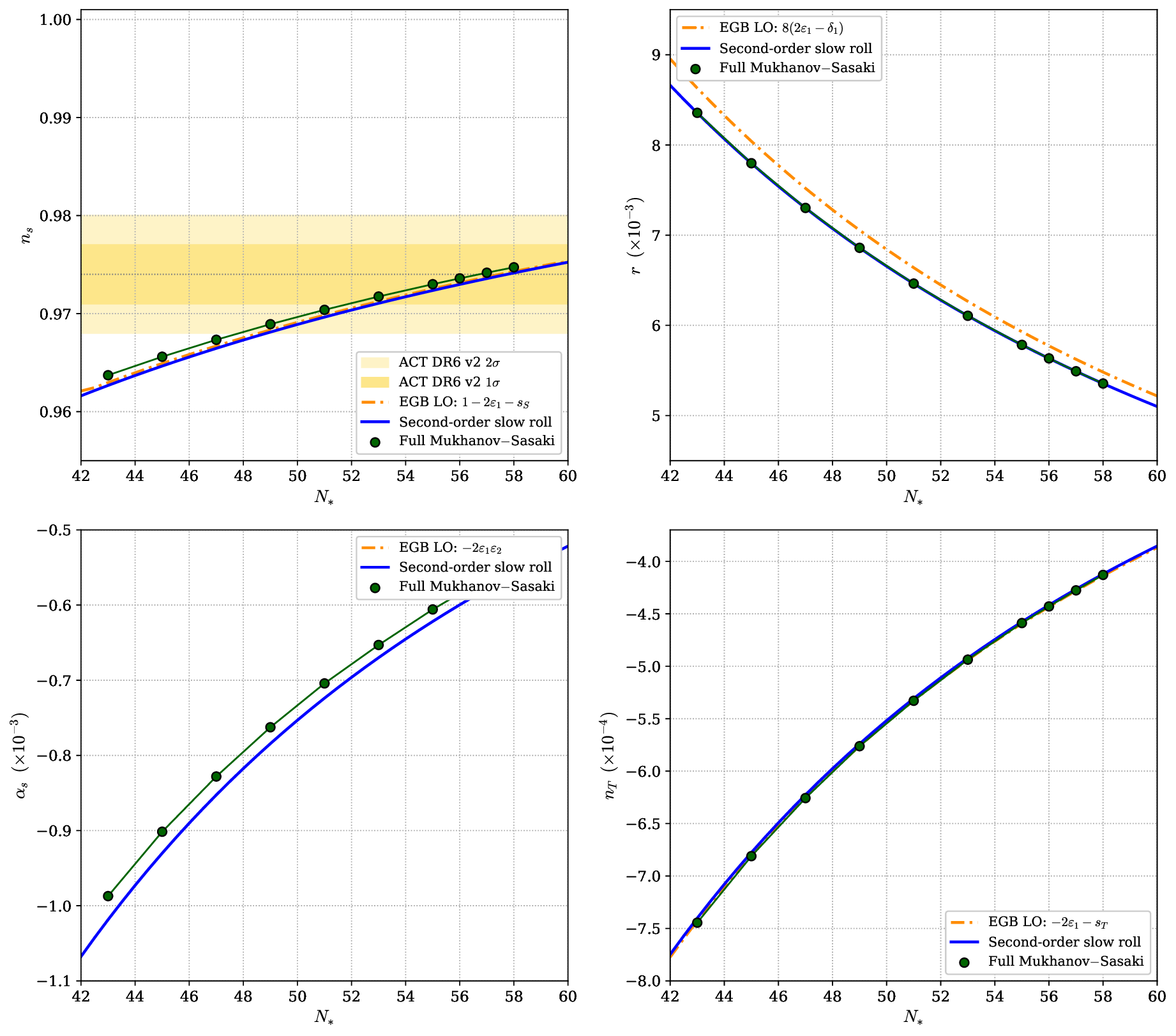}
\caption{Comparison of three computational methods on $N_{*}\in[42,60]$. Orange dash-dotted: EGB LO consistency relations (Sec.~\ref{subsec:model_methods}, (i)). Blue solid: 2nd-order phase-$\theta$ closure of this work (the conventional $(\phi,\dot\phi,H)$ scheme of~\cite{Koh:2016abf} collapses to the same blue line). Dark-green markers and connecting line: full Mukhanov--Sasaki integration. Panels: $n_{s}$ (top left, with ACT~DR6 v2 $1\sigma$/$2\sigma$ bands), $r$ (top right), $\alpha_{s}$ (bottom left), $n_{T}$ (bottom right).}
\label{fig:model_methods}
\end{figure}

\begin{figure}[H]
\centering
\includegraphics[width=0.95\textwidth]{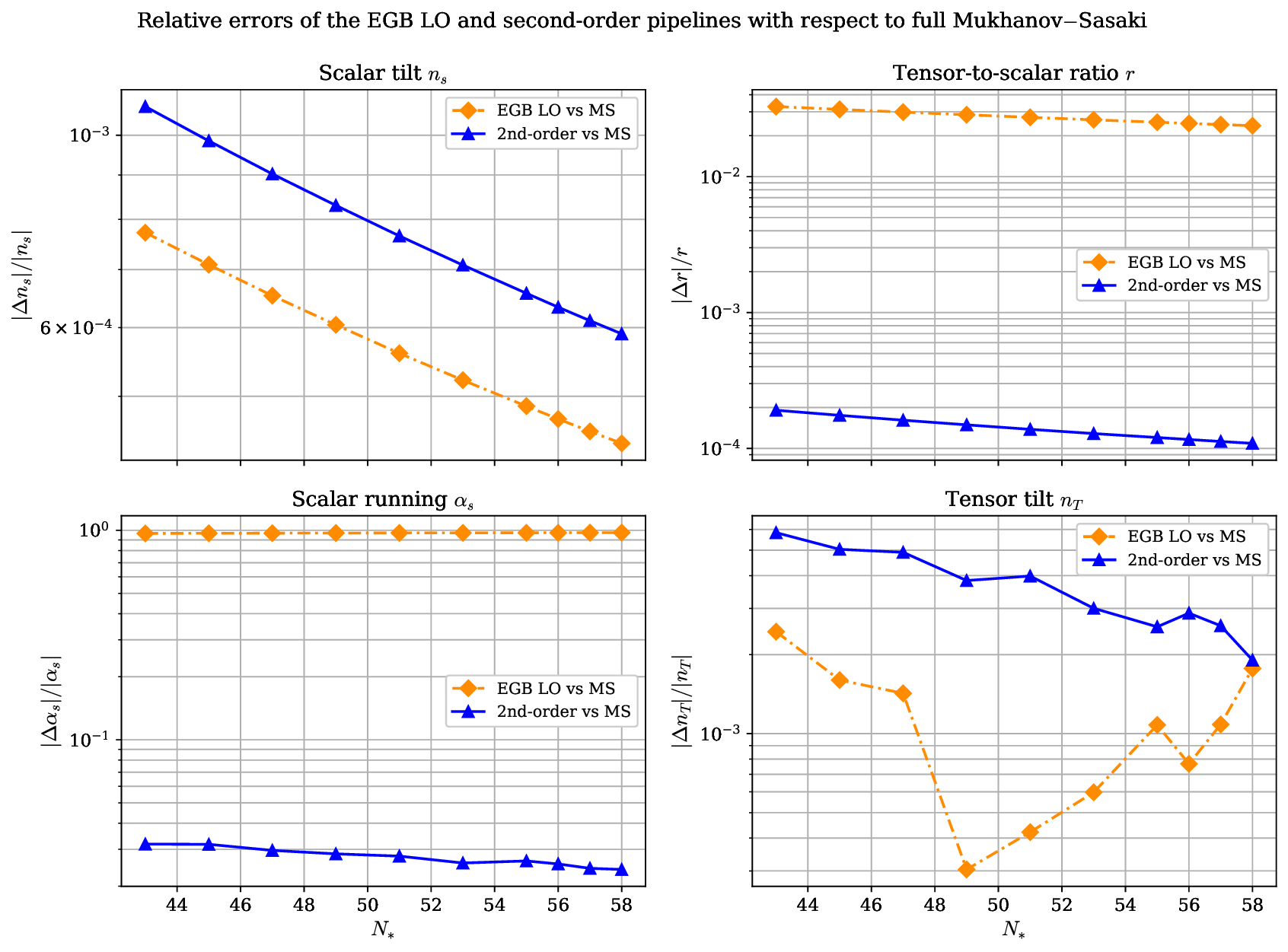}
\caption{Relative errors $|\Delta X|/|X|$ of the EGB LO consistency relations (red markers, dash-dotted) and of the 2nd-order phase-$\theta$ closure (blue squares, solid) with respect to the full Mukhanov--Sasaki integration, for $n_{s}, r, \alpha_{s}, n_{T}$ on $N_{*}\in[43,58]$. For $\alpha_{s}$ the LO truncation is $\mathcal{O}(1)$ because the second-order Hankel pieces dominate the running (cf.\ Sec.~\ref{subsec:model_methods}).}
\label{fig:methods_relerr}
\end{figure}

We compare three levels of computation for $n_{s}, r, \alpha_{s}, n_{T}$ as functions of the pivot e-fold $N_{*}$:
\begin{enumerate}[label=(\roman*)]
\item \textbf{EGB leading-order} (LO) slow-roll consistency relations
\begin{align*}
n_{s,\rm LO}-1 &= -2\eps_{1}-s_{S}, & s_{S} &= \frac{2\eps_{1}\eps_{2}-\delta_{1}(\delta_{2}-\eps_{1})}{2\eps_{1}-\delta_{1}},\\
n_{T,\rm LO}   &= -2\eps_{1}-s_{T}, & s_{T} &= \Delta(\eps_{1}-\delta_{2}),\\
r_{\rm LO}     &= 8(2\eps_{1}-\delta_{1}), & \alpha_{s,\rm LO} &= -2\eps_{1}\eps_{2},
\end{align*}
each defined as the leading non-trivial slow-roll truncation of the corresponding second-order expression~\eqref{eq:ns_2nd}--\eqref{eq:alpha_def} that retains every $\mathcal{O}(\mathrm{SR})$ piece (so the GB-induced $s_{S}, s_{T}$ contributions and the GB-corrected $r$ prefactor enter at LO already), and evaluated along the inflationary trajectory at horizon crossing. The strictly first-order-in-$\eps_{1}$ truncation $n_{s}-1\simeq -2\eps_{1}$, used in earlier non-EGB analyses, drops the equally-LO $s_{S}$ piece and would inflate the LO residual on $n_{s}$ at the pivot from $\sim 5\times 10^{-4}$ to $\sim 2.7\times 10^{-2}$ (a factor $\sim 50$); the EGB LO above is what we plot in Figs.~\ref{fig:model_methods}--\ref{fig:methods_relerr}.
\item \textbf{Phase-$\theta$ second-order} (this work) --- the closed-form chain of Sec.~\ref{subsec:model_closed} assembled into the Hankel-index expressions $\nu_{S}, \nu_{T}$ of~\eqref{eq:nu_indices} and substituted into the exact prefactor form~\eqref{eq:r_exact} of the tensor-to-scalar ratio. The auxiliary derivatives $s_{A}, \beta_{A}$ and $s_{c_{A}}, \beta_{c_{A}}$ are obtained by fourth-order finite differences of $\ln Q_{A}$ and $\ln c_{A}$ along the trajectory.
\item \textbf{Full Mukhanov--Sasaki integration} (MS), as described in Sec.~\ref{subsec:model_method}, with $n_{s}, r, n_{T}, \alpha_{s}$ extracted by quadratic regression of $\ln\Delta_{A}^{2}(k)$ in $\ln k$ around $k_{*}$.
\end{enumerate}
The phase-$\theta$ pipeline is computed entirely through the closed-form expressions of Secs.~\ref{sec:theta} and~\ref{subsec:model_closed}, with no slow-roll truncation other than the second-order Hankel-index expansion of Appendix~\ref{app:derivation}. The MS pipeline is the standard mode-by-mode integration described in Sec.~\ref{subsec:model_method}.

The pivot values at $N_{*}=55$ for each method are tabulated quantity-by-quantity in Table~\ref{tab:closed_vs_MS}. Across the full scan window $N_{*}\in[42,60]$, the analytical phase-$\theta$ closure of this work (computed symbolically from the $(\theta,H)$ chain) agrees with our independent numerical implementation of the same second-order slow-roll formalism in the conventional $(\phi,\dot{\phi},H)$ state vector (which follows the analytical scheme of Refs.~\cite{Hwang:2005hb,Koh:2014bka,Koh:2016abf,Koh:2018qcy}) to five significant digits. This is an internal cross-check that the analytical $(\theta,H)$ pipeline reproduces the result obtained by following the conventional scheme. The gap between the second-order curves and the full MS markers, visible on the $n_{s}$ and $r$ panels of Fig.~\ref{fig:model_methods}, is dominated by the $\sim 7\%$ Friedmann-constraint drift on $\delta_{2}/s_{S}$ documented later in this subsection and by the finite MS-regression precision; the strict analytical Hankel truncation contributes only $\mathcal{O}(\mathrm{SR}^{3})\lesssim 10^{-6}$ and is sub-dominant. All numerical sources sit well below the current ACT~DR6 v2 measurement uncertainty $\sigma_{n_{s}}=3.0\times10^{-3}$.

Quantitatively, the relative errors of the EGB LO and the second-order pipelines with respect to the full Mukhanov--Sasaki integration, $\delta X\equiv|X-X_{\rm MS}|/|X_{\rm MS}|$ for $X\in\{n_{s},r,\alpha_{s},n_{T}\}$, are computed across the full MS scan window $N_{*}\in[43,58]$ and shown in Fig.~\ref{fig:methods_relerr}. At the pivot $N_{*}=55$ the EGB LO formulae of Sec.~\ref{subsec:model_methods}~(i) give $\delta n_{s}\simeq 5\times 10^{-4}$, $\delta r\simeq 2.5\times 10^{-2}$, $\delta n_{T}\simeq 1.1\times 10^{-3}$, and a much larger $\delta\alpha_{s}\simeq 0.97$: the running is the one observable for which the leading slow-roll piece $-2\eps_{1}\eps_{2}$ is too small in absolute value because the next-order pieces $-\beta_{S}-\beta_{c_{S}}$ dominate the full~\eqref{eq:alpha_def}, so the LO truncation has no practical accuracy for $\alpha_{s}$ and must be replaced by the second-order expression. The second-order phase-$\theta$ closure of this work brings the residual on $r$ down to $\delta r\sim 1.2\times 10^{-4}$ at the pivot, with $\delta n_{s}\sim 6\times 10^{-4}$, well below the ACT~DR6 v2 measurement uncertainty on $n_{s}$ and the BICEP/Keck upper bound on $r$; for $n_{s}$ and $n_{T}$ the corrected EGB LO is already accurate at this level for the present realization with $|\delta_{1}|\sim 3\times 10^{-4}$, so the decisive gain of the second order is in $r$ and $\alpha_{s}$. For the running $\alpha_{s}$ the second-order residual is at the percent level ($\delta\alpha_{s}\sim 2.6\times 10^{-2}$ at the pivot), corresponding to an absolute deviation $|\Delta\alpha_{s}|\sim 1.6\times 10^{-5}$, still well inside the ACT~DR6 measurement uncertainty $\sigma_{\alpha_{s}}=5.2\times 10^{-3}$ (i.e.~$\sigma_{\alpha_{s}}/|\alpha_{s}|\sim 10$). For the tensor tilt $n_{T}$ the second-order residual stays in the $10^{-3}$ regime, comparable to the LO.

Table~\ref{tab:closed_vs_MS} lists, at the pivot $N_{*}=55$, every intermediate quantity produced by the closed-form chain of Sec.~\ref{subsec:model_closed} side by side with the values obtained from the full Mukhanov--Sasaki integration. The two pipelines agree to five significant digits on $\nu_{S}, \nu_{T}, n_{s}, n_{T}, r$ and on every coefficient of the perturbation action; the residual $\sim 7\%$ on $\delta_{2}$ and $s_{S}$ reflects the $\sim10^{-4}$ Friedmann-constraint drift of the dynamic integrator amplified through the near-plateau inversion $V_{0}(1-e^{-\sqrt{2/3}\phi})^{2}\to\phi$, and is invisible at the level of the observable spectral parameters.

\begin{table}[H]
\centering
\small
\begin{threeparttable}
\caption[Closed-form vs. full Mukhanov--Sasaki at the pivot]{Quantity-by-quantity comparison of the closed-form pipeline of Sec.~\ref{subsec:model_closed} with the full Mukhanov--Sasaki integration at the pivot $N_{*}=55$, $(\theta_{*},H_{*})=(0.011085,\,7.7498\times10^{-6}\,\Mpl)$.}\label{tab:closed_vs_MS}
\begin{tabular}{@{}lccc@{}}
\toprule
Quantity & Closed-form (this work) & Full MS integration & Relative deviation\\
\midrule
$\eps_{1}$               & $2.278\times10^{-4}$ & $2.278\times10^{-4}$ & $<10^{-5}$\\
$\eps_{2}$               & $3.579\times10^{-2}$ & $3.579\times10^{-2}$ & $<10^{-4}$\\
$|\delta_{1}|$           & $2.86\times10^{-4}$  & $2.86\times10^{-4}$  & $<10^{-4}$\\
$|\delta_{2}|$           & $1.24\times10^{-2}$  & $1.33\times10^{-2}$  & $7\%$\\
$Q_{T}$                  & $1.00029$            & $1.00029$            & $<10^{-5}$\\
$Q_{S}$                  & $7.374\times10^{-4}$ & $7.374\times10^{-4}$ & $<10^{-5}$\\
$c_{T}$                  & $0.99986$            & $0.99986$            & $<10^{-5}$\\
$c_{S}$                  & $1.0000$             & $1.0000$             & $<10^{-5}$\\
$\nu_{T}$                & $1.50023$            & $1.50023$\tnote{a}    & $<10^{-5}$\\
$\nu_{S}$                & $1.51382$            & $1.5135$\tnote{a}     & $2\times 10^{-4}$\\
$n_{T}$                  & $-4.58\times10^{-4}$ & $-4.58\times10^{-4}$ & $<10^{-3}$\\
$n_{s}$                  & $0.9724$             & $0.9730$             & $6\times10^{-4}$\\
$r$                      & $5.783\times10^{-3}$ & $5.784\times10^{-3}$ & $2\times10^{-4}$\\
\bottomrule
\end{tabular}
\begin{tablenotes}\footnotesize
\item[a] The MS pipeline does not extract $\nu_{S},\nu_{T}$ directly; the values listed are $(4-n_{s}^{\rm MS})/2$ and $(3-n_{T}^{\rm MS})/2$ respectively, obtained from the MS-regression spectral tilts via the Hankel-index relations $n_{s}-1=3-2\nu_{S}$ and $n_{T}=3-2\nu_{T}$.
\end{tablenotes}
\end{threeparttable}
\end{table}

For the observable parameters $n_{s}$ and $r$ the agreement is well below the ACT~DR6 v2 measurement uncertainty $\sigma_{n_{s}}=3.0\times10^{-3}$, demonstrating that the analytic chain~\eqref{eq:dxi_starlin}--\eqref{eq:sS_starlin} reproduces the full numerical computation at observational accuracy for this realization while keeping every step transparent. The running of the scalar spectral index $\alpha_{s}=-2\eps_{1}\eps_{2}-\beta_{S}-\beta_{c_{S}}$ involves second derivatives along the trajectory and is more sensitive to the integrator drift; the LO MS regression value $\alpha_{s}=-6.06\times10^{-4}$ remains compatible with ACT~DR6 ($\sigma_{\alpha_{s}}=5.2\times10^{-3}$) and is what we quote in Table~\ref{tab:model_obs}.

\section{Relic gravitational-wave spectrum}\label{sec:omega_gw}

Relic, or primordial, gravitational waves are the transverse-traceless tensor perturbations of the metric that are amplified out of the Bunch--Davies vacuum during inflation~\cite{DeFelice:2010aj}. Each tensor mode is frozen on super-horizon scales during inflation, re-enters the comoving Hubble radius later in the radiation- or matter-dominated era and propagates from then on as a free transverse-traceless metric ripple, contributing to the present-day stochastic gravitational-wave background~\cite{Grishchuk:2005qe,Boyle:2005se,Watanabe:2006qe,Caprini:2018mtu}. Because gravitons decouple from the cosmic fluid below the Planck scale, the relic background propagates essentially unscathed through every subsequent epoch and so encodes the entire post-inflationary expansion history, including the equation of state of the reheating era and the effective number of relativistic species~\cite{Koh:2018qcy,Nakayama:2008wy,Cook:2015vqa}. The shape of the relic spectrum simultaneously probes (i) the inflationary energy scale through the amplitude $\Delta_{t}^{2}(k_{*})\propto H_{*}^{2}$, (ii) the deviation from the canonical de Sitter consistency relation through the tensor tilt $n_{T}$ and its running $\alpha_{T}$, and (iii) the reheating dynamics through the location of the spectral break $f_{\rh}$. For the EGB models considered here this background therefore provides a complementary probe to the CMB observables: $n_{s}$ and $r$ at the pivot constrain the inflationary plateau, while the high-frequency tail of $\Omega_{\GW}(f)$ tests the dynamics in the end-of-inflation region $\theta\to\theta_{e}$ and the subsequent reheating phase. Comprehensive reviews of cosmological gravitational-wave backgrounds and of their detection programmes can be found, for example, in~\cite{Grishchuk:2005qe,Boyle:2005se,Watanabe:2006qe,Caprini:2018mtu}.

\begin{figure}[H]
\centering
\includegraphics[width=0.95\textwidth]{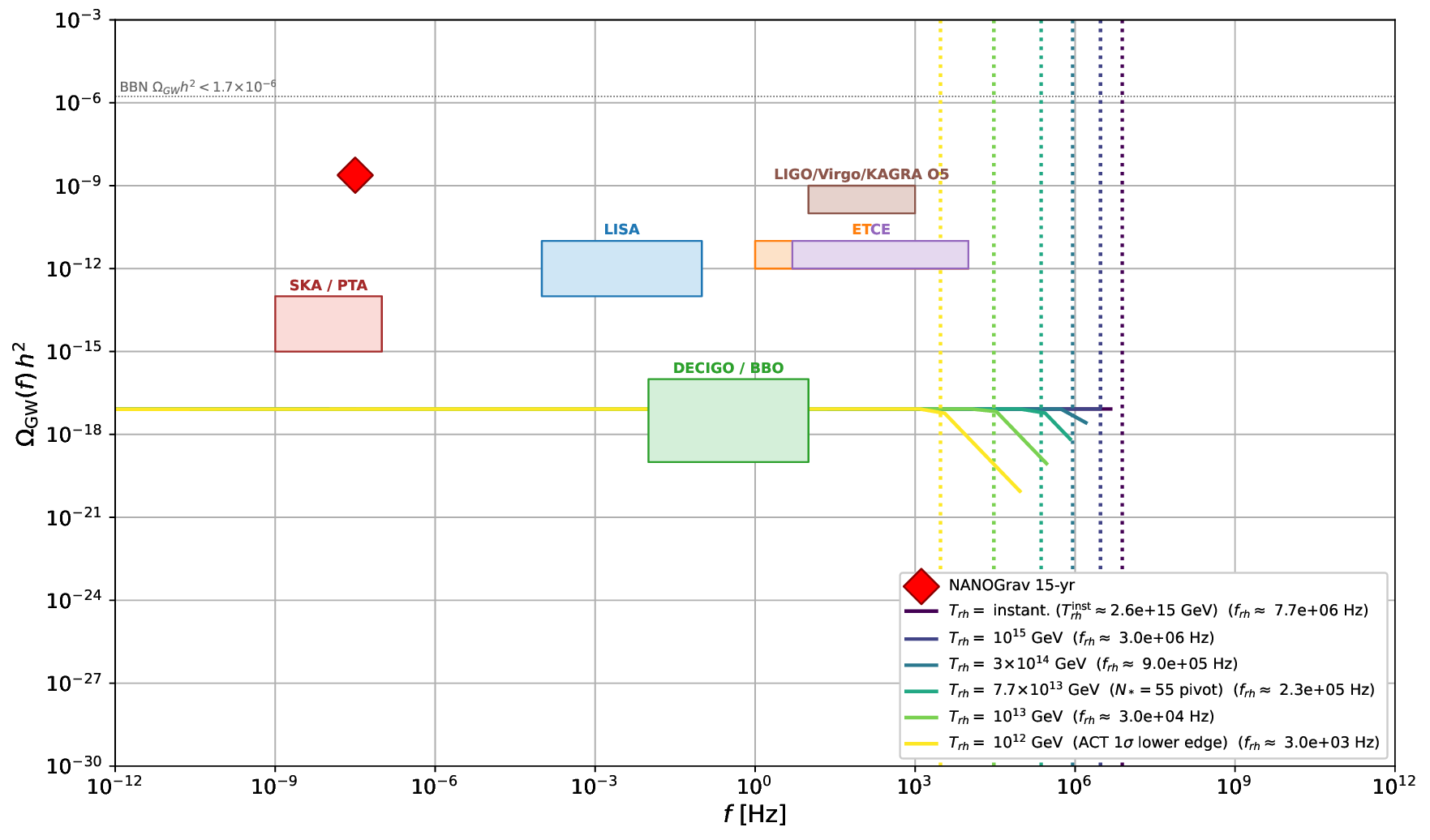}
\caption{Relic gravitational-wave spectrum $\Omega_{\GW}(f)\,h^{2}$ predicted by the model~\eqref{eq:specific_model} for six values of $T_{\rm rh}\in[10^{12},\,2.56\times10^{15}]~\mathrm{GeV}$ inside the ACT~DR6 $1\sigma$ window for $n_{s}$. Coloured dotted lines: spectral-break frequencies $f_{\rm rh}$. Shaded bands: published PLS sensitivity intervals of SKA PTA, LISA, DECIGO/BBO, ET, CE and LIGO/Virgo/KAGRA O5 (Table~\ref{tab:detector_PLS}). Spectra above the upper band edge exceed the detection threshold; spectra below the lower edge are buried in noise. Red diamond: NANOGrav 15-year measurement~\cite{NANOGrav:2023gor} at $f_{\rm yr}\approx 3\times 10^{-8}~\mathrm{Hz}$. Gray dotted: BBN bound $\Omega_{\GW}h^{2}<1.7\times 10^{-6}$.}
\label{fig:model_OmegaGW}
\end{figure}

The dimensionless energy density of the relic gravitational-wave background today reads~\cite{Boyle:2005se,Watanabe:2006qe,Caprini:2018mtu}
\begin{equation}
\Omega_{\GW}(f)h^{2}=\frac{1}{12}\Delta_{t}^{2}(f)\,\mathcal{T}^{2}(f),
\label{eq:Omega_GW}
\end{equation}
with $\Delta_{t}^{2}(f)$ the primordial tensor spectrum and $\mathcal{T}(f)$ the transfer function across the post-inflationary epochs. We extrapolate the spectrum from the pivot using the standard quadratic-logarithmic form~\cite{Kosowsky:1995aa}
\begin{equation}
\Delta_{t}^{2}(f)=\Delta_{t}^{2}(k_{*})\,\exp\!\left[\!\left(n_{T}+\tfrac{1}{2}\alpha_{T}\ln(f/f_{*})\right)\!\ln(f/f_{*})\right],
\label{eq:Dt2_extrap}
\end{equation}
where $f=k/(2\pi a_{0})$ at adiabatic expansion from the pivot, $f_{*}\approx7.7\times10^{-17}~\mathrm{Hz}$. The transfer function is computed in the idealized two-stage model: instantaneous radiation domination at $f<f_{\rh}$ and the matter-dominated scaling $\Omega_{\GW}\propto f^{-2}$ at $f>f_{\rh}$. The $\Omega_{\GW}\propto f^{-2}$ scaling in the matter-like reheating window is the $\bar w_{\rm rh}\to 0$ limit of the general $f^{2(3\bar w_{\rm rh}-1)/(3\bar w_{\rm rh}+1)}$ behaviour of~\cite{Koh:2018qcy,Nakayama:2008wy,Cook:2015vqa}; the studied model value $\bar w_{\rm rh}=+0.0025\ll 1$ shifts the spectral exponent by less than $2\%$ and is absorbed into the precision of Fig.~\ref{fig:model_OmegaGW}.

The detector bands in Fig.~\ref{fig:model_OmegaGW} display the power-law-integrated sensitivity (PLS) $\Omega^{\rm PLS}_{\GW}(f)\,h^{2}$ of each facility~\cite{Thrane:2013oya}. The PLS represents the smallest amplitude of a stochastic power-law background that produces a signal-to-noise ratio above the chosen detection threshold once the SNR is integrated over the full operating band of the (cross-correlated) detector pair~\cite{Caprini:2018mtu,Thrane:2013oya}. Explicitly, it is the envelope over all power-law slopes $\beta$ of the single-slope detection thresholds~\cite{Thrane:2013oya},
\begin{equation}
\Omega^{\rm PLS}_{\GW}(f)\;=\;\max_{\beta}\!\left[\,\Omega_{\beta}\!\left(\frac{f}{f_{\rm ref}}\right)^{\!\beta}\,\right]\!,\qquad
\Omega_{\beta}\;=\;\frac{\mathrm{SNR}_{\rm thr}}{\sqrt{2T_{\rm obs}}}\,\!\left[\,\int_{f_{\rm lo}}^{f_{\rm hi}}\!\!df\;\frac{(f/f_{\rm ref})^{2\beta}}{\Omega_{\rm eff}^{2}(f)}\,\right]^{\!-1/2}\!\!,
\label{eq:PLS_def}
\end{equation}
with $T_{\rm obs}$ the observation time, $\mathrm{SNR}_{\rm thr}$ the chosen detection threshold (taken as $\mathrm{SNR}_{\rm thr}=1$ by~\cite{Thrane:2013oya}), and $\Omega_{\rm eff}(f)\equiv(10\pi^{2}/3H_{0}^{2})f^{3}S_{n}(f)/\gamma(f)$ the effective noise energy density built from the strain noise power spectral density $S_{n}(f)$ and the overlap reduction function $\gamma(f)$ of the cross-correlated detector pair. The PLS is the appropriate detection threshold for a smooth cosmological background such as~\eqref{eq:Dt2_extrap}, and the canonical sensitivity envelope used in recent inflationary-GW analyses~\cite{Oikonomou:2023bli,Caprini:2018mtu,Vagnozzi:2023lwo}. Each shaded band spans the facility's operating frequency interval and the published PLS amplitude interval; the values used here are summarized with their literature sources in Table~\ref{tab:detector_PLS}.

\begin{table}[H]
\centering
\caption{Power-law-integrated sensitivity (PLS) intervals $\Omega^{\rm PLS}_{\GW}(f)\,h^{2}$ used in Fig.~\ref{fig:model_OmegaGW}, with literature sources. PLS convention and references: Sec.~\ref{sec:omega_gw}.}\label{tab:detector_PLS}
\begin{tabular}{@{}lcccc@{}}
\toprule
Facility & $[f_{\rm lo},f_{\rm hi}]$ (Hz) & $\Omega^{\rm PLS}_{\rm min}\,h^{2}$ & $\Omega^{\rm PLS}_{\rm max}\,h^{2}$ & Source\\
\midrule
SKA PTA              & $10^{-9}$--$10^{-7}$ & $10^{-15}$ & $10^{-13}$ & \cite{Caprini:2018mtu,Janssen:2014dka}\\
LISA                 & $10^{-4}$--$10^{-1}$ & $10^{-13}$ & $10^{-11}$ & \cite{LISA:2017pwj,Robson:2018ifk}\\
DECIGO / BBO         & $10^{-2}$--$10$      & $10^{-19}$ & $10^{-16}$ & \cite{Corbin:2005ny,Kawamura:2020pcg}\\
ET                   & $1$--$10^{4}$        & $10^{-12}$ & $10^{-11}$ & \cite{Hild:2010id}\\
CE                   & $5$--$10^{4}$        & $10^{-12}$ & $10^{-11}$ & \cite{Reitze:2019iox}\\
LIGO/Virgo/KAGRA O5  & $10$--$10^{3}$       & $10^{-10}$ & $10^{-9}$  & \cite{LIGOScientific:2014pky,KAGRA:2021kbb}\\
\bottomrule
\end{tabular}
\end{table}

The plotted frequency window in Fig.~\ref{fig:model_OmegaGW} starts at $f_{\rm min}=10^{-12}~\mathrm{Hz}$. This lower bound is dictated by two complementary considerations. First, the pulsar-timing-array (PTA) and ultra-low-frequency probes that anchor the present-day relic GW phenomenology operate above this scale: square-kilometre-array PTA observations span $\sim 10^{-16}$--$10^{-9}~\mathrm{Hz}$~\cite{NANOGrav:2023gor} and the NANOGrav 15-year detection lies in the nHz band, with reference frequency $f_{\rm yr}\approx 3\times 10^{-8}~\mathrm{Hz}$. Below $\sim 10^{-15}~\mathrm{Hz}$ the spectrum is constrained by the CMB tensor amplitude itself~\cite{BICEP:2021xfz,Caprini:2018mtu,Planck:2018jri} rather than by a direct GW measurement, so the same physical content is already encoded in the $(n_{s},r)$ constraint we use in Sec.~\ref{subsec:model_cmb}. Second, the cosmological transfer function in the standard two-stage description~\cite{Boyle:2005se,Watanabe:2006qe,Nakayama:2008wy}
crosses over from the matter-era plateau to the radiation-era scaling at the frequency of horizon re-entry at matter--radiation equality, $f_{\rm eq}\sim 1.6\times 10^{-17}~\mathrm{Hz}$~\cite{Watanabe:2006qe}; restricting the plot to $f\geq 10^{-12}~\mathrm{Hz}$ stays well inside the radiation-era branch of the transfer function and avoids the additional contamination from the matter-era $f^{-2}$ knee. The window $f\in[10^{-12},10^{8}]~\mathrm{Hz}$ therefore covers, in a single panel, all the operating and projected GW facilities relevant to the discussion (PTA/SKA, LISA, DECIGO/BBO, LIGO/Virgo/KAGRA, ET/CE, and the higher-frequency electromagnetic-conversion programmes~\cite{Aggarwal:2020olq,Domcke:2022rgu,Bringmann:2023gba}).

The result is shown in Fig.~\ref{fig:model_OmegaGW} for six reheating temperatures in the window $T_{\rm rh}\in[10^{12}~\mathrm{GeV},\,T_{\rm rh}^{\rm inst}]$, where the lower bound corresponds to the ACT~DR6 v2 $1\sigma$ requirement $n_{s}\geq0.971$ and the upper bound to instantaneous reheating. The peak value of $\Omega_{\GW}(f)h^{2}$ remains below $\sim 10^{-17}$ over the entire observable frequency range, eight orders of magnitude below the NANOGrav signal $\Omega_{\GW}^{\rm NG}h^{2}\approx2.4\times10^{-9}$. This is consistent with the general conclusion drawn in the literature that EGB inflation in the standard slow-roll regime cannot account for PTA-band signals without additional amplification mechanisms such as sourced or second-order tensor production, non-attractor enhancement, or scalar-induced gravitational waves~\cite{Koh:2018qcy,Oikonomou:2023bli,Vagnozzi:2023lwo,Franciolini:2023pbf}.

The position of the spectral break $f_{\rh}$ ranges from $f_{\rh}\sim 3\times 10^{3}~\mathrm{Hz}$ at the ACT~DR6 $1\sigma$ lower edge ($T_{\rm rh}\sim 10^{12}~\mathrm{GeV}$) to $f_{\rh}\approx 7.7\times 10^{6}~\mathrm{Hz}$ at instantaneous reheating, with the central, physically preferred reheating range ($T_{\rm rh}\sim 10^{13}$--$10^{15}~\mathrm{GeV}$) sitting at $f_{\rh}\sim 10^{4}$--$10^{6}~\mathrm{Hz}$. The break itself lies well above the operating bands of all currently planned detectors, identifying the ultra-high-frequency window as the strategic longer-term target for testing the EGB-specific reheating signature of this class of models. The plateau amplitude $\Omega_{\GW}h^{2}\sim 10^{-17}$--$10^{-18}$ across the decihertz window, on the other hand, lies above the PLS sensitivity floor of full DECIGO and BBO (Table~\ref{tab:detector_PLS}) and therefore constitutes a target signal for that family of space-based interferometers (full DECIGO; the B-DECIGO precursor configuration would not see this signal because its strain noise is roughly two orders of magnitude higher across the decihertz band~\cite{Kawamura:2020pcg}).

\section{Discussion}\label{sec:discussion}

The phase-$\theta$ parametrization developed here brings two methodological gains. First, the inflationary observables in the EGB framework are reduced to explicit closed-form functions of $(H,\theta,\dot{\xi},\ddot{\xi})$ for arbitrary potential $V(\phi)$ and arbitrary non-minimal coupling $\xi(\phi)$. The exact form~\eqref{eq:r_param} of the tensor-to-scalar ratio is, to our knowledge, the simplest fully analytic expression available in the EGB setup that smoothly interpolates between the deep slow-roll regime and the end of inflation, where the standard $r=16\eps_{1}$ relation is most strongly modified. Second, the parametrization remains regular through the entire inflationary trajectory, including the kinematic singularity $\delta_{1}=1$ of the tensor coefficient $Q_{T}=1-\delta_{1}$: the phase $\theta$ varies monotonically from $\theta\to0$ at the onset of inflation to the end-of-inflation value $\theta_{e}$ defined by $\eps_{1}(\theta_{e})=1$ (in the GR limit $\sin^{2}\theta_{e}=1/3$, with a small Gauss--Bonnet correction in the general case), providing a stable independent variable for both analytical and numerical work.

On the post-inflationary side, the same parametrization yields an analytical handle on the reheating equation of state $\bar w_{\rm rh}$: starting from the exact Friedmann pair~\eqref{eq:EGB_F1}--\eqref{eq:EGB_F2} one obtains $w_{\rm eff}$ in closed form (\ref{eq:w_eff_theta}), and the time average over the coherent oscillations around the quadratic Starobinsky minimum vanishes at leading order~\cite{Turner:1983he}, guaranteeing $\bar w_{\rm rh}\to 0^{+}>-1/3$ analytically; the numerical integration confirms $\bar w_{\rm rh}=+0.0025$.

The application to the Starobinsky inflation with linear EGB-coupling function shows that the second-order analytical formulae match the full Mukhanov--Sasaki integration with an accuracy comfortably better than the ACT~DR6 measurement uncertainty on $n_{s}$, with explicit relative residuals $|\Delta n_{s}|/n_{s}\sim 6\times 10^{-4}$ and $|\Delta r|/r\sim 1.2\times 10^{-4}$ at the pivot (Fig.~\ref{fig:methods_relerr}). This makes the phase-$\theta$ formulae a practical tool for rapid scans of model space against the updated CMB constraints. The realization itself sits inside the ACT~DR6 $1\sigma$ window for $n_{s}$, well below the BICEP/Keck upper bound on $r$, in compliance with the inflationary-epoch CMB B-mode bound on $c_{T}$ and with the late-universe GW170817 bound on $c_{T}$ (both secured as discussed in Sec.~\ref{subsec:model_bg}, item~(3) of the viability checks), and with the post-inflationary Gauss--Bonnet positivity bound $N_{\rm GB}\leq 2 N_{E}$ of Ref.~\cite{Fomin:2020hfh}. The closed-form pipeline gives $n_{s}=0.9724$ while the full Mukhanov--Sasaki integration gives $n_{s}=0.9730$ (Table~\ref{tab:closed_vs_MS}): the headline number we quote in Table~\ref{tab:model_obs} is the MS value; the $6\times 10^{-4}$ residual is dominated by the $\sim 7\%$ Friedmann-constraint drift on $\delta_{2}/s_{S}$ documented in Sec.~\ref{subsec:model_methods} and by the finite MS-regression precision; the strict analytical Hankel truncation contributes only $\mathcal{O}(\mathrm{SR}^{3})\lesssim 10^{-6}$ and is negligible (Appendix~\ref{app:derivation}, A.6). All numerical sources sit well inside the ACT~DR6 v2 $1\sigma$ uncertainty $\sigma_{n_{s}}=3.0\times 10^{-3}$.

On the gravitational-wave side, the predicted plateau amplitude $\Omega_{\GW}(f)h^{2}\sim 10^{-17}$--$10^{-18}$ exceeds the published PLS detection threshold of full DECIGO and BBO ($\Omega^{\rm PLS}_{\GW}h^{2}\sim 10^{-19}$, Table~\ref{tab:detector_PLS}~\cite{Thrane:2013oya,Corbin:2005ny,Kawamura:2020pcg}) and is therefore observable as a stochastic signal by that family of decihertz space interferometers, even though it remains below the PLS thresholds of SKA PTA, LISA, ET, CE and LIGO/Virgo/KAGRA. The NANOGrav signal at $\Omega_{\GW}^{\rm NG}h^{2}\sim 2\times10^{-9}$ remains eight orders of magnitude above the predicted spectrum and would require an additional amplification mechanism (sourced gravitational waves, second-order tensors from enhanced scalar perturbations, or a blue tensor tilt from a non-attractor phase), which is absent in the minimal realization studied here. The EGB-induced spectral break $f_{\rh}$ falls in the ultra-high-frequency band, $f_{\rh}\sim10^{4}$--$10^{6}~\mathrm{Hz}$ for the physically preferred reheating window $T_{\rm rh}\sim 10^{13}$--$10^{15}~\mathrm{GeV}$, currently inaccessible but a natural longer-term target for next-generation electromagnetic-conversion experiments and axion haloscopes~\cite{Aggarwal:2020olq,Domcke:2022rgu,Bringmann:2023gba}.

\section{Conclusion}\label{sec:conclusion}

We have developed a phase-$\theta$ parametrization for inflation in Einstein--Gauss--Bonnet gravity that reduces the geometric slow-roll parameters, the coefficients $Q_{T}, Q_{S}$ of the quadratic action of scalar and tensor perturbations, the propagation speeds, the Hankel indices, the dimensionless power spectra, the tensor-to-scalar ratio and the spectral tilts and their runnings to explicit closed-form functions of the Hubble rate, the phase $\theta$ and the derivatives of the non-minimal coupling $\xi(\phi)$. The exact expression $r=r(H,\theta,\dot{\xi},\ddot{\xi})$ furnishes a transparent analytic bridge between fundamental Lagrangian parameters and observable quantities, valid through the entire inflationary stage.

Applied to a Starobinsky potential complemented by a minimal linear Gauss--Bonnet coupling $\xi(\phi)=\xi_{0}\phi$, the formalism yields $n_{s}=0.9730$, $r=5.78\times10^{-3}$, $\alpha_{s}=-6.06\times10^{-4}$ at the pivot $N_{*}=55$, consistent with ACT~DR6 v2 + BICEP/Keck and with the Gauss--Bonnet positivity bound $N_{\rm GB}\leq 2 N_{E}$. The inflationary and late-universe constraints on $c_{T}$ are both satisfied, as discussed in Sec.~\ref{subsec:model_bg}, item~(3) of the viability checks. The second-order analytical formulae differ from the full Mukhanov--Sasaki integration by $|\Delta n_{s}|\approx 6\times10^{-4}$, a factor of $\sim 5$ below the ACT~DR6 v2 uncertainty $\sigma_{n_{s}}=3.0\times10^{-3}$, with the residual reflecting the $\sim 7\%$ Friedmann-constraint drift on $\delta_{2}/s_{S}$ and the finite MS-regression precision; the strict analytical truncation $\mathcal{O}(\mathrm{SR}^{3})\lesssim 10^{-6}$ is sub-dominant (cf.\ Sec.~\ref{subsec:model_methods}).

The associated relic gravitational-wave spectrum, $\Omega_{\GW}(f)h^{2}\sim 10^{-17}$--$10^{-18}$ on the inflationary plateau, exceeds the power-law-integrated detection threshold of full DECIGO and BBO and is therefore observable as a stochastic signal by decihertz space interferometers, while remaining below the PLS thresholds of SKA PTA, LISA, ET, CE and LIGO/Virgo/KAGRA O5 (Table~\ref{tab:detector_PLS}). The EGB-induced spectral break $f_{\rh}$ sits in the ultra-high-frequency band, $f_{\rh}\sim 10^{4}$--$10^{6}~\mathrm{Hz}$ for the physically preferred reheating window, identifying that band as the natural longer-term target for tests of this class of models.

Finally, it should be noted that the advancement of high-precision measurement techniques in modern cosmology, including forthcoming gravitational-wave detectors, necessitates the corresponding development of more accurate methods for analysing cosmological models. Accordingly, a natural extension of the proposed approach lies in its application and generalization to inflationary models based on other types of modified theories of gravity~\cite{Sotiriou:2008rp,DeFelice:2011bh,DeFelice:2011zh,DeFelice:2011jm,
Capozziello:2011et,Ishak:2018his,CANTATA:2021asi,Bahamonde:2021gfp,Odintsov:2023weg}.

\section*{Acknowledgements}

The work of I.V.F. was carried out within the framework of Supplementary Agreement No. 073-03-2026-035/1 dated 21.02.2026 to the Agreement on the provision of a subsidy to a federal budgetary or autonomous institution for financial support for the implementation of a state assignment for the provision of public services (performance of work) No. 073-03-2026-035 dated 23.01.2026, concluded between the Federal State Budgetary Educational Institution of Higher Education ``UlGPU named after I.N. Ulyanov" and the Ministry of Education of the Russian Federation.

%
%
%
%
%
%

\appendix
\section{Detailed derivation of the second-order observables}\label{app:derivation}

This appendix collects the technical steps used to derive the second-order slow-roll expansion of the Hankel indices~\eqref{eq:nu_indices} and the closed-form expressions for $r$, $n_{s}$, $n_{T}$, $\alpha_{s}$, $\alpha_{T}$ used in the main text. The presentation does not claim new physical results: its individual ingredients are due to Hwang and Noh~\cite{Hwang:2005hb} (universal quadratic action), Guo and co-workers~\cite{Guo:2009uk,Guo:2010jr}, Satoh~\cite{Satoh:2010ep}, Kuroyanagi--Chiba--Sugiyama~\cite{Kuroyanagi:2010mm}, Wu, Zhu and Wang~\cite{Wu:2017joj}, and Koh and collaborators~\cite{Koh:2014bka,Koh:2016abf,Koh:2018qcy} (adaptation to EGB) and Ref.~\cite{Kaur:2023wos} (phase-$\theta$ parametrization for canonical single-field general-relativistic inflation, here extended to the EGB sector). Its purpose is to compile them into a single self-contained derivation in which every step is explicitly controlled and compared with the GR limit.

\subsection{Exact form of \texorpdfstring{$z_{A}''/z_{A}$}{}}

For an arbitrary function $f$ one has $f'=a\dot{f}$. Define $g_{A}\equiv\dot{z}_{A}/z_{A}$ and the reduced parameters
\begin{equation}
s_{A}\equiv\frac{\dot{Q}_{A}}{HQ_{A}},\qquad \beta_{A}\equiv\frac{\dot{s}_{A}}{H}.
\label{eq:sA_betaA}
\end{equation}
From $z_{A}^{2}=a^{2}Q_{A}$,
\begin{equation}
g_{A}=H+\tfrac{1}{2}\frac{\dot{Q}_{A}}{Q_{A}}=H\!\left(1+\tfrac{s_{A}}{2}\right)\!,
\label{eq:gA}
\end{equation}
and differentiation yields the exact identity
\begin{equation}
\frac{z_{A}''}{z_{A}}=a^{2}H^{2}\!\left[(1+s_{A}/2)(2-\eps_{1}+s_{A}/2)+\tfrac{\beta_{A}}{2}\right]\!.
\label{eq:zApp_exact}
\end{equation}

For approximately constant $\eps_{1}$ (power-law inflation), $\mathcal{H}\eta=-1/(1-\eps_{1})+\mathcal{O}(\eps_{1}\eps_{2})$, so
\begin{equation}
\eta^{2}\frac{z_{A}''}{z_{A}}=\frac{(1+s_{A}/2)(2-\eps_{1}+s_{A}/2)+\beta_{A}/2}{(1-\eps_{1})^{2}}+\mathcal{O}(\text{SR}^{3}).
\label{eq:eta2_zApp}
\end{equation}

\subsection{Second-order expansion of \texorpdfstring{$\nu_{A}$}{}}

Substituting~\eqref{eq:eta2_zApp} into $\nu_{A}^{2}=1/4+\eta^{2}z_{A}''/z_{A}$ and expanding to second order in slow roll,
\begin{align}
\nu_{A}^{2}&=\tfrac{9}{4}+3\eps_{1}+\tfrac{3s_{A}}{2}+4\eps_{1}^{2}+\tfrac{5}{2}s_{A}\eps_{1}+\tfrac{s_{A}^{2}}{4}+\tfrac{\beta_{A}}{2}+\mathcal{O}(\text{SR}^{3}),\label{eq:nuA2_exp}\\
\nu_{A}&=\tfrac{3}{2}+\eps_{1}+\tfrac{s_{A}}{2}+\eps_{1}^{2}+\tfrac{s_{A}\eps_{1}}{2}+\tfrac{\beta_{A}}{6}+\mathcal{O}(\text{SR}^{3}).\label{eq:nuA_2nd_app}
\end{align}
For variable $c_{A}$, the change of variable $d\tilde{\eta}_{A}=c_{A}d\eta$ replaces $s_{A}\to s_{A}+s_{c_{A}}$ with $s_{c_{A}}=\dot{c}_{A}/(Hc_{A})$, reproducing~\eqref{eq:nu_indices}.

\subsection{Bessel solution and Bunch--Davies vacuum}

The general solution of~\eqref{eq:Bessel} is
\begin{equation}
v_{A}(\eta)=\sqrt{-\eta}\!\left[C_{1}H_{\nu_{A}}^{(1)}(-c_{A}k\eta)+C_{2}H_{\nu_{A}}^{(2)}(-c_{A}k\eta)\right]\!,
\label{eq:general_sol}
\end{equation}
where $H_{\nu}^{(1,2)}=J_{\nu}\pm iY_{\nu}$. Bunch--Davies asymptotics, $v_{A}\to e^{-ic_{A}k\eta}/\sqrt{2c_{A}k}$ for $-c_{A}k\eta\to\infty$, fix $C_{1}=\sqrt{\pi}/2\,e^{i\pi(\nu_{A}+1/2)/2}$, $C_{2}=0$, and the super-horizon limit of the Hankel function $H_{\nu}^{(1)}(y\to0)\to-(i/\pi)\Gamma(\nu)(y/2)^{-\nu}$ together with the reflection formula $\Gamma(\nu)\Gamma(1-\nu)=\pi/\sin(\pi\nu)$ produces
\begin{equation}
|v_{A}(\eta,k)|^{2}=\frac{\csc^{2}(\pi\nu_{A})}{4\pi\Gamma^{2}(1-\nu_{A})}(-\eta)\!\left(\frac{c_{A}k(-\eta)}{2}\right)^{\!-2\nu_{A}}\!\!.
\label{eq:v_square}
\end{equation}

Evaluating at sound-horizon crossing $c_{A}k=aH$ ($-\eta\simeq1/(aH)$ at leading order), $k^{3}(-\eta)/a^{2}=H^{2}/c_{S}^{3}$ for the scalar mode and analogously for the tensor mode, one recovers the dimensionless spectra~\eqref{eq:PS_full}--\eqref{eq:PT_full}.

\subsection{Slow-roll parameters \texorpdfstring{$s_{T}$, $s_{S}$}{} in phase-\texorpdfstring{$\theta$}{} variables}

From $Q_{T}=1-\delta_{1}$ and $\dot{\delta}_{1}/H=\delta_{1}(\delta_{2}-\eps_{1})$, one obtains $s_{T}=\Delta(\eps_{1}-\delta_{2})$. Using~\eqref{eq:eps1_theta},
\begin{equation}
s_{T}=\Delta\!\left[3\sin^{2}\theta+\tfrac{1}{2}\Delta(1-\delta_{2})-\delta_{2}\right]\!.
\label{eq:sT_theta_app}
\end{equation}
For the scalar mode, $Q_{S}\simeq 2\eps_{1}-\delta_{1}$ in our convention $\delta_{1}\equiv 4H\dot{\xi}$, and the slow-roll derivatives, using $\dot{\eps}_{1}/H=\eps_{1}\eps_{2}$ and $\dot{\delta}_{1}/H=\delta_{1}(\delta_{2}-\eps_{1})$, yield
\begin{equation}
s_{S}=\frac{2\eps_{1}\eps_{2}-\delta_{1}(\delta_{2}-\eps_{1})}{2\eps_{1}-\delta_{1}}.
\label{eq:sS_theta_app}
\end{equation}
The second slow-roll parameter $\eps_{2}$ is obtained from differentiation of~\eqref{eq:eps1_theta} using~\eqref{eq:Hdot_theta}, \eqref{eq:thetadot_theta} together with the explicit form of $\xi(\phi)$.

\subsection{ Tensor-to-scalar ratio in phase-\texorpdfstring{$\theta$}{} variables}

Taking the ratio of~\eqref{eq:PT_full} and~\eqref{eq:PS_full},
\begin{equation}
r=8\,\frac{\Gamma^{2}(1-\nu_{S})\csc^{2}(\pi\nu_{T})}{\Gamma^{2}(1-\nu_{T})\csc^{2}(\pi\nu_{S})}\,2^{2(\nu_{T}-\nu_{S})}\,\frac{Q_{S}c_{S}^{3}}{Q_{T}c_{T}^{3}}\bigg|_{*}.
\label{eq:r_exact_app}
\end{equation}
At first order in slow roll, $\nu_{S}-\nu_{T}=(s_{S}-s_{T})/2$, the gamma- and cosecant-functions contribute only at second order, and~\eqref{eq:r_exact_app} reduces to~\eqref{eq:r_LO}. Substitution of~\eqref{eq:QT_theta}--\eqref{eq:QScS_SR} and resummation of the leading-order corrections in $(\dot{\xi},\ddot{\xi})$ yields the compact closed form~\eqref{eq:r_param}.

\subsection{Summary of observables}

Combining the results,
\begin{align}
\delta_{1}&=4H\dot{\xi},\quad \delta_{2}=\frac{\ddot{\xi}}{H\dot{\xi}},\quad \Delta=\frac{\delta_{1}}{1-\delta_{1}},\\
\eps_{1}(\theta)&=3\sin^{2}\theta+\tfrac{1}{2}\Delta(1-\delta_{2}),\\
Q_{T}(\theta)&=1-\delta_{1},\quad c_{T}^{2}(\theta)=1-\Delta(\delta_{2}-1),\\
Q_{S}(\theta)&\simeq 2\eps_{1}-\delta_{1}+\mathcal{O}(\mathrm{SR}^{2}),\quad c_{S}^{2}(\theta)\simeq 1,\\
s_{T}&=\Delta(\eps_{1}-\delta_{2}),\quad s_{S}=\frac{2\eps_{1}\eps_{2}-\delta_{1}(\delta_{2}-\eps_{1})}{2\eps_{1}-\delta_{1}},\\
\Delta_{s}^{2}(k)&=\frac{\csc^{2}(\pi\nu_{S})\,2^{2\nu_{S}-3}}{\pi\Gamma^{2}(1-\nu_{S})}\frac{H^{2}}{Q_{S}c_{S}^{3}}\bigg|_{c_{S}k=aH},\\
\Delta_{t}^{2}(k)&=\frac{\csc^{2}(\pi\nu_{T})\,2^{2\nu_{T}}}{\pi\Gamma^{2}(1-\nu_{T})}\frac{H^{2}}{Q_{T}c_{T}^{3}}\bigg|_{c_{T}k=aH},\\
r(\theta)&=8\!\left[6\sin^{2}\theta-\frac{4(\ddot{\xi}-4H^{2}\dot{\xi}^{2})}{1-4H\dot{\xi}}\right]\!,\\
n_{s}-1&\simeq-2\eps_{1}-s_{S}-s_{c_{S}}-2\eps_{1}^{2}-s_{S}\eps_{1}-\tfrac{1}{3}\beta_{S},\\
n_{T}&\simeq-2\eps_{1}-s_{T}-s_{c_{T}}-2\eps_{1}^{2}-s_{T}\eps_{1}-\tfrac{1}{3}\beta_{T},\\
\alpha_{s}&\simeq-2\eps_{1}\eps_{2}-\beta_{S}-\beta_{c_{S}},\quad \alpha_{T}\simeq-2\eps_{1}\eps_{2}-\beta_{T}-\beta_{c_{T}}.
\end{align}
For the model~\eqref{eq:specific_model} with $|\delta_{1}|\sim3\times10^{-4}$, the strict next-order analytical correction in slow roll is $\mathcal{O}(\text{SR}^{3})\lesssim10^{-6}$, so the residuals quoted in Sec.~\ref{subsec:model_methods} are dominated by the integrator and regression precision rather than by the analytical truncation.

\bibliographystyle{JHEP}
\bibliography{references2}

\end{document}